\documentclass[aps,twocolumn,showpacs,amsmath,amssymb,superscriptaddress,prc]{revtex4-2}
\usepackage{graphicx,here}
\usepackage{multirow}
\usepackage{tikz}
\usepackage{epstopdf}
\usepackage[percent]{overpic}
\usepackage{scrextend}
\usepackage{xcolor,xspace}
\usepackage[ulem=normalem]{changes}
\usepackage{hyperref}
\hypersetup{colorlinks=True,urlcolor=blue,linkcolor=blue,citecolor=blue,filecolor=black}

\colorlet{Changes@Color}{red}
\usepackage{dcolumn,color,footnote,bm,braket}
\usepackage{url,longtable,tabularx}
\usepackage{threeparttable}

\usepackage{fancybox}
\usetikzlibrary{matrix}

\newcommand\+{\dagger}
\newcommand\jr{j_{\rho}}
\newcommand\mr{m_{\rho}}

\newcommand\mgt{M_{\mathrm{GT}}}

\newcommand\hb{\hat{H}_{\mathrm{B}}}
\newcommand\hf{\hat{H}_{\mathrm{F}}}
\newcommand\hbf{\hat{V}_{\mathrm{BF}}}
\newcommand\hff{\hat{V}_{\nu\pi}}

\newcommand\ga{g_{\mathrm{A}}}
\newcommand\gv{g_{\mathrm{V}}}

\newcommand\ft{\log{}_{10}ft}
\newcommand\btm{\beta^{-}}

\newcommand\vd{v_{\mathrm{d}}}

\newcommand\vsss{v_{\mathrm{ss}}}
\newcommand\vt{v_{\mathrm{t}}}

\begin{document}

\title{Parameter dependence of the $\beta$-decay properties of neutron-rich Zr isotopes within the interacting boson model}

\author{M. Homma}
\affiliation{Department of Physics, 
Hokkaido University, Sapporo 060-0810, Japan}

\author{K. Nomura}
\email{nomura@sci.hokudai.ac.jp}
\affiliation{Department of Physics, 
Hokkaido University, Sapporo 060-0810, Japan}
\affiliation{Nuclear Reaction Data Center, 
Hokkaido University, Sapporo 060-0810, Japan}

\date{\today}

\begin{abstract}
We investigate parameter dependence 
of the calculated $\beta$-decay 
properties, as well as low-lying states for the neutron-rich 
Zr isotopes within 
the neutron-proton interacting boson model (IBM-2) and 
interacting boson-fermion-fermion model (IBFFM-2). 
It is shown that the calculated $\log_{10}ft$ values for the 
transitions of the $0^+_1$ ground states of the 
parent even-even nuclei $^{96-102}$Zr into the $1^+_1$ 
states of the daughter odd-odd nuclei $^{96-102}$Nb 
consistently exhibit a strong dependence on those parameters 
associated with the quadrupole-quadrupole boson interaction, 
and with the residual interaction between 
an unpaired neutron and an unpaired proton in the IBFFM-2 
Hamiltonian for the odd-odd Nb nuclei. 
By the reduction in magnitude of the quadrupole-quadrupole 
interaction strength by approximately a factor of 2, 
the calculated 
$\log_{10}ft$ values for the Zr$(0^+_1)\to$Nb$(1^+_1)$ 
transitions increase and agree with the experimental values. 
This points to a significant improvement over 
the previous study performed in the same mass region, 
that consists of the mapping from 
a relativistic energy density functional calculation 
onto the IBM-2 Hamiltonian. 
\end{abstract}

\maketitle

\section{Introduction}

Nuclear $\beta$-decay is a process in which a 
neutron in a nucleus is converted into 
a proton, or vice versa, 
emitting an electron (positron) 
and an antielectron (electron) neutrino. 
The $\beta$ decay plays an essential role in the 
rapid neutron-capture processes in astrophysical 
nucleosynthesis, which produce heavy chemical elements, 
and is also used as an experimental technique 
to measure energy levels of a given nucleus. 
The $\beta$-decay rates of numerous neutron-rich 
heavy nuclei have been measured extensively by experiments 
in major radioactive-ion-beam facilities 
worldwide 
\cite{dillmann2003,nishimura2011,quinn2012,lorusso2015,caballero2016}, 
which also calls for reliable theoretical 
predictions.

Accurate theoretical predictions as well as 
experimental measurements of the $\beta$ decay also 
provide input to evaluate the double-$\beta$ decay 
nuclear matrix elements 
\cite{avignone2008,engel2017,agostini2023}. 
It is a rare process 
in which, when the $Q_{\beta}$ value of the 
single-$\beta$ decay from the even-even to odd-odd 
nuclei is high enough, a decay process may 
occur between neighboring even-even nuclei with the 
neutron and proton numbers ($N,Z$) = ($N\mp2,Z\pm2$), 
emitting two electrons (positrons) and some 
light particles. 
Theoretical evaluation of the double-$\beta$ decay nuclear 
matrix elements currently differs by a factor of 2-3 
among different nuclear models, and reduction of the 
theoretical uncertainty is under active investigation.

Precise calculations of the nuclear 
wave functions for the low-lying states of the 
initial and final nuclei, 
which enter the relevant transition operators, 
are crucial to provide reliable predictions of 
the $\beta$-decay properties, such as the half-lives 
and $\ft$ values. 
Theoretical approaches that allow for the calculation of 
the $\beta$-decay properties along with a quantitative 
and detailed description of the low-energy excitations 
have been made, e.g., by the nuclear shell model 
\cite{langanke2003,caurier2005,syoshida2018,suzuki2018}, 
quasiparticle random phase approximations 
\cite{alvarez2004,sarriguren2015,boillos2015,pirinen2015,
simkovic2013,mustonen2016,suhonen2017,ravlic2021,yoshida2023}, 
and the interacting boson model (IBM) 
\cite{navratil1988,dellagiacoma1988phdthesis,
DELLAGIACOMA1989,brant2004,brant2006,yoshida2013,
mardones2016,nomura2020beta-1,nomura2020beta-2,ferretti2020,
nomura2022beta-ge,nomura2022beta-rh,Vsevolodovna2022,nomura2024beta}.

In a recent article \cite{nomura2024beta}, 
a simultaneous description of the low-lying states 
and (allowed) $\beta$-decay properties 
of the neutron-rich even-even and odd-odd 
nuclei from the $_{36}$Kr to $_{48}$Cd isotopes 
near $N=60$ was made by using 
the IBM that is based on the 
energy density functional (EDF) framework. 
In Ref.~\cite{nomura2024beta}, 
the constrained self-consistent mean-field 
(SCMF) calculations were performed within 
the relativistic Hartree-Bogoliubov (RHB) model 
\cite{vretenar2005,niksic2011}
using the density-dependent point-coupling (DD-PC1) 
EDF \cite{DDPC1} and the separable pairing 
force of finite range \cite{tian2009}. 
The constrained calculations provided 
each even-even nucleus with 
the potential energy surface (PES) in terms 
of the triaxial quadrupole deformations, 
which is then mapped onto the expectation value 
of the neutron-proton IBM (IBM-2) 
Hamiltonian in the boson condensate state 
\cite{ginocchio1980} for the even-even nuclei. 
This mapping procedure completely determines 
the parameters of the IBM-2 Hamiltonian. 
The same RHB SCMF calculation produced 
single-particle energies and 
occupation probabilities at spherical configuration 
for the neighboring odd-odd 
nuclei, and these quantities were 
used to determine the neutron-proton interacting 
boson-fermion-fermion model 
(IBFFM-2) \cite{brant1984,IBFM} Hamiltonian for the odd-odd systems. 
Remaining coupling constants for the 
interactions between an odd neutron and 
an odd proton, and between an odd nucleon 
and even-even boson core were fit to 
reproduce a few low-lying 
states of each odd-odd nucleus 
to a reasonable accuracy. 
The Gamow-Teller (GT) transition strengths between 
an even-even and an odd-odd nuclei have been 
computed by using those nuclear wave functions 
that were obtained from the IBM-2 and IBFFM-2 
Hamiltonians, respectively, and were 
employed for calculating the $\beta$-decay 
$\ft$ values.

It was found in Ref.~\cite{nomura2024beta}, 
however, that the $\ft$ values 
for the $\btm$ decay of the studied even-even 
into odd-odd nuclei, especially for 
those of Kr, Sr, and Zr nuclei near $N=60$ and $Z=40$, 
are systematically underestimated by a factor 
of $\approx 1.5$ within the mapped IBM-2 framework. 
These too small $\ft$ values, hence too large GT 
transition matrix elements, imply 
some deficiencies of this method, which 
could be effectively accounted for 
by introducing certain 
amount of quenching of the axial-vector coupling 
constant $\ga$. 
The significant discrepancy between the experimental 
and calculated $\ft$ values encountered in 
Ref.~\cite{nomura2024beta}, however, would require 
one to use an unrealistically large quenching factor 
to reproduce the data. 
A possible cause of this discrepancy was 
also investigated in that study, and 
was attributed to the choice of the employed EDF 
that is a basis for determining most of the 
model parameters. 
Another possible factors that could give rise 
to the too small $\ft$ values would lie in the 
IBM-2 and IBFFM-2 calculations, which provide 
nuclear wave functions for the parent and daughter 
nuclei.

In the present article, we report an extensive 
analysis of the dependence 
of the $\ft$ predictions on various model parameters 
involved in the nuclear structure calculations within 
the IBM-2 and IBFFM-2, and attempt to 
identify which of the model parameters affect 
most significantly the final results 
on the $\ft$ values. 
We show systematic behaviors 
of the calculated properties, such 
as the excitation energies for each 
parent and daughter nuclei, and the $\ft$ values, 
as functions of a given model parameter. 
The present study is based on the previous work 
of Ref.~\cite{nomura2024beta}, and hence 
the Hamiltonian parameters, and other ingredients, 
e.g., single-particle energies and occupation 
probabilities, that were obtained in the mapped IBM-2 
calculations of Ref.~\cite{nomura2024beta} are 
here considered as a starting point for our analysis. 
In that way, we explore 
possible ways of improving the mapped 
IBM-2 description of $\beta$ decay. 
We further calculate low-lying states of 
each of the parent and daughter nuclei to investigate 
how an optimal set of parameters that 
gives a reasonable agreement with the observed 
$\ft$ values can reproduce the experimental energy 
spectrum in comparison to the previous calculation 
\cite{nomura2024beta}. 
The parameter dependence of the excitation energies 
of odd-odd nuclei has been investigated within 
the IBFFM-2, e.g., in Ref.~\cite{Vsevolodovna2022}, 
but to the best of our knowledge the behaviors 
of the $\ft$ values according to the 
IBFFM-2 parameters have not been studied.

Considering the fact that a large number of 
independent parameters enter the model, 
which in principle correlate with others and also 
differ from one nucleus to another, to keep the 
discussion as simple as possible we here focus on 
the $\btm$ decay of the $^{96-102}$Zr isotopes.  
The low-lying structure of the neutron-rich Zr 
isotopes has attracted considerable attention, since 
they exhibit a rapid shape transition at 
$N\approx 60$ and 
a possible shape coexistence \cite{heyde2011}. 
$^{96}$Zr is also a candidate nucleus for 
the two-neutrino and neutrinoless double-$\beta$ 
decays, and its nuclear structural and single-$\beta$ 
decay properties should be 
of much relevance.

The paper is structured as follows. 
In Sec.~\ref{sec:model} we review the IBM-2 and IBFFM-2 
Hamiltonians, their parameters, and GT transition 
operators, which were used in Ref.~\cite{nomura2024beta}. 
We show in Sec.~\ref{sec:ft-para} 
the parameter dependence of the $\ft$ 
values on the model parameters. 
The variations with the parameters of the 
calculated excitation energies are discussed 
in Sec.~\ref{sec:energy-para}. 
Section~\ref{sec:summary} gives a summary, 
and perspectives for a future study.

\section{Theoretical framework\label{sec:model}}

This section gives a brief reminder of the 
IBM-2 and IBFFM-2 frameworks for calculating 
the $\beta$ decay matrix elements. 
More detailed descriptions of these 
theoretical framework  
are found in Ref.~\cite{nomura2024beta} 
for the neutron-rich Zr region, and for 
general descriptions of the $\beta$-decay studies 
of odd-mass and odd-odd nuclei the reader is referred to 
Refs.~\cite{DELLAGIACOMA1989,dellagiacoma1988phdthesis,IBFM} 
and \cite{brant2006,yoshida2013}, respectively.

\subsection{Model Hamiltonian and GT operator}

Within the IBM-2, an even-even core nucleus 
is described in terms of the neutron $s_{\nu}$ and 
$d_{\nu}$ bosons, and the proton $s_{\pi}$ and 
$d_{\pi}$ bosons \cite{IBM}. 
From a microscopic point of view, the neutron 
(proton) $s_{\nu}$ and $d_{\nu}$ 
($s_{\pi}$ and $d_{\pi}$) bosons represent 
collective monopole and quadrupole pairs of 
valence neutrons (protons), respectively \cite{IBM,OAIT,OAI}. 
For the IBM-2 Hamiltonian we take the form
\begin{align}
\label{eq:bham}
 \hb = 
&\epsilon_{d}(\hat{n}_{d_{\nu}}+\hat{n}_{d_{\pi}})
+\kappa\hat{Q}_\nu \cdot\hat{Q}_\pi
\nonumber\\
&+\kappa_{\nu}\hat{Q}_\nu \cdot\hat{Q}_\nu
+\kappa_{\pi}\hat{Q}_\pi \cdot\hat{Q}_\pi
+ \kappa'\hat{L}\cdot\hat{L} \; ,
\end{align}
where the first term 
stands for the $d$-boson number operator with 
$\hat n_{d_\rho} = d^\+_\rho\cdot\tilde d_\rho$ 
($\rho=\nu$ or $\pi$) and with $\epsilon_d$ 
the single $d$ boson energy.  
The second, third, and fourth terms are 
the quadrupole-quadrupole 
interactions between neutron and proton bosons, 
between neutron and neutron bosons, 
and between proton and proton bosons, 
respectively. 
The quadrupole operator $\hat Q_{\rho}$ is 
defined as 
$\hat Q_{\rho} = s^\+_\rho \tilde d_\rho + d^\+ s_{\rho} + \chi_{\rho} (d^\+_\rho\times\tilde d_\rho)^{(2)}$, 
with $\chi_\nu$ and $\chi_\pi$ being
dimensionless parameters. 
$\kappa$, $\kappa_{\nu}$, and $\kappa_\pi$ 
are strength parameters. 
The fifth term in Eq.~(\ref{eq:bham}) 
stands for a rotational term, 
with $\kappa'$ being 
the strength parameter, and 
$\hat L = \hat L_{\nu} + \hat L_{\pi}$
denotes the angular momentum operator 
with $\hat L_\rho = (d^\+_\rho\times\tilde d^\+_\rho)^{(1)}$.

The IBM-2 is extended to treat 
odd-odd nuclear systems by including 
an unpaired neutron and an 
unpaired proton degrees of freedom, and their 
couplings. 
The IBFFM-2 Hamiltonian is expressed in general by
\begin{align}
\label{eq:ham}
 \hat{H}=\hb + \hf^{\nu} + \hf^{\pi} + \hbf^{\nu} + \hbf^{\pi} + \hff \; .
\end{align}
The first term represents the IBM-2 core 
Hamiltonian (\ref{eq:bham}). 
The second and third terms of 
Eq.~(\ref{eq:ham}) represent 
the single-neutron and -proton Hamiltonians, 
respectively, and take the form
\begin{align}
\label{eq:hf}
 \hf^{\rho} = -\sum_{\jr}\epsilon_{\jr}\sqrt{2\jr+1}
  (a_{\jr}^\+\times\tilde a_{\jr})^{(0)}
\equiv
\sum_{\jr}\epsilon_{\jr}\hat{n}_{\jr} \; ,
\end{align}
where $\epsilon_{\jr}$ stands for the 
single-particle energy of the odd neutron 
or proton orbital $\jr$. 
$a_{\jr}^{(\+)}$ represents 
the particle annihilation (creation) operator, 
with $\tilde{a}_{\jr}$ defined by 
$\tilde{a}_{\jr\mr}=(-1)^{\jr -\mr}a_{\jr-\mr}$. 
On the right-hand side of Eq.~(\ref{eq:hf}), 
$\hat{n}_{\jr}$ stands for the number operator 
for the odd particle. 
The single-particle space taken in the 
present study comprises the neutron $3s_{1/2}$, 
$2d_{3/2}$, $2d_{5/2}$, and $1g_{7/2}$ orbitals, 
and the proton $1g_{9/2}$ orbital in the 
$N=50-82$ and $Z=28-50$ 
major oscillator shells for calculating 
the positive-parity states of the 
odd-odd Nb nuclei. 
For $^{96-102}$Nb, since the odd neutron and odd proton 
are, respectively, treated as a particle 
and a hole, the corresponding even-even 
boson cores are the $^{96-102}$Mo nuclei, 
respectively.

The fourth (fifth) term on the 
right-hand side  
of Eq.~(\ref{eq:ham}) denotes the interaction 
between a single neutron (or proton) and 
the even-even boson core, and is 
given as \cite{scholten1985,IBFM}
\begin{equation}
\label{eq:hbf}
 \hbf^{\rho}
=\Gamma_{\rho}\hat{V}_{\mathrm{dyn}}^{\rho}
+\Lambda_{\rho}\hat{V}_{\mathrm{exc}}^{\rho}
+A_{\rho}\hat{V}_{\mathrm{mon}}^{\rho} \; ,
\end{equation}
where the first, second, and third terms 
represent the quadrupole dynamical, exchange, 
and monopole interactions, respectively, 
with the strength parameters 
$\Gamma_\rho$, $\Lambda_\rho$, and $A_{\rho}$. 
Each term in the above expression reads
\begin{widetext}
\begin{align}
\label{eq:dyn}
&\hat{V}_{\mathrm{dyn}}^{\rho}
=\sum_{\jr\jr'}\gamma_{\jr\jr'}
(a^{\+}_{\jr}\times\tilde{a}_{\jr'})^{(2)}
\cdot\hat{Q}_{\rho'},\\
\label{eq:exc}
&\hat{V}^{\rho}_{\mathrm{exc}}
=-\left(
s_{\rho'}^\+\times\tilde{d}_{\rho'}
\right)^{(2)}
\cdot
%\left(
\sum_{\jr\jr'\jr''}
\sqrt{\frac{10}{N_{\rho}(2\jr+1)}}
\beta_{\jr\jr'}\beta_{\jr''\jr}
:\left[
(d_{\rho}^{\+}\times\tilde{a}_{\jr''})^{(\jr)}\times
(a_{\jr'}^{\+}\times\tilde{s}_{\rho})^{(\jr')}
\right]^{(2)}:
+ (\text{H.c.}) \; ,\\
\label{eq:mon}
&\hat{V}_{\mathrm{mon}}^{\rho}
=\hat{n}_{d_{\rho}}\hat{n}_{\jr} \; ,
\end{align}
\end{widetext}
where the $j$-dependent factors 
$\gamma_{\jr\jr'}=(u_{\jr}u_{\jr'}-v_{\jr}v_{\jr'})Q_{\jr\jr'}$, 
and $\beta_{\jr\jr'}=(u_{\jr}v_{\jr'}+v_{\jr}u_{\jr'})Q_{\jr\jr'}$, 
with 
$Q_{\jr\jr'}=\braket{\ell_{\rho}\frac{1}{2}\jr\|Y^{(2)}\|\ell'_\rho\frac{1}{2}\jr'}$ being the matrix element of the fermion 
quadrupole operator in the single-particle basis. 
$\hat{Q}_{\rho'}$ in Eq.~(\ref{eq:dyn}) denotes 
the quadrupole operator in the boson system, 
introduced in Eq.~(\ref{eq:bham}). 
The notation $:(\cdots):$ in Eq.~(\ref{eq:exc}) 
stands for normal ordering. 
Note that the forms of $\hbf^{\rho}$ have been 
discussed on microscopic grounds 
in Refs.~\cite{scholten1985,IBFM}. 
Within this framework 
the unperturbed single-particle 
energy, $\epsilon_{\jr}$, in Eq.~(\ref{eq:hf}) 
should be replaced with the quasiparticle 
energy $\tilde\epsilon_{\jr}$.

The last term of Eq.~(\ref{eq:ham}), $\hff$, 
corresponds to the residual interaction 
between the unpaired neutron and proton. 
The following form is here considered 
for this interaction. 
\begin{align}
\label{eq:hff}
\hff
=& 4\pi{\vd}
\delta(\bm{r})
\delta(\bm{r}_{\nu}-r_0)
\delta(\bm{r}_{\pi}-r_0)
\nonumber
\\
&+ \vt
\left[
\frac{3({\bm\sigma}_{\nu}\cdot{\bf r})
({\bm\sigma}_{\pi}\cdot{\bf r})}{r^2}
-{\bm{\sigma}}_{\nu}
\cdot{\bm{\sigma}}_{\pi}
\right] \; .
\end{align}
The first and second terms stand for the $\delta$, 
and tensor interactions, with 
$\vd$, and $\vt$ being strength parameters, 
respectively. 
Note that $\bm{r}=\bm{r}_{\nu}-\bm{r}_{\pi}$ 
is the relative coordinate of the 
neutron and proton, and $r_0=1.2A^{1/3}$ fm. 
The matrix element of $\hff$ depends on the 
occupation $v_j$ and unoccupation $u_j$ 
amplitudes.

The GT transition operator is here defined by 
\begin{align}
\label{eq:ogt}
&\hat{T}^{\rm GT}
=\sum_{j_{\nu}j_{\pi}}
\eta_{j_{\nu}j_{\pi}}^{\mathrm{GT}}
\left(\hat P_{j_{\nu}}\times\hat P_{j_{\pi}}\right)^{(1)} \; ,
\end{align}
with the coefficients $\eta$ calculated as
\begin{align}
\label{eq:etagt}
\eta_{j_{\nu}j_{\pi}}^{\mathrm{GT}}
&= - \frac{1}{\sqrt{3}}
\left\langle
\ell_{\nu}\frac{1}{2}j_{\nu}
\bigg\|{\bm\sigma}\bigg\|
\ell_{\pi}\frac{1}{2}j_{\pi}
\right\rangle
\delta_{\ell_{\nu}\ell_{\pi}} \; .
\end{align}
$\hat P_{j_\nu}$ and $\hat P_{j_\pi}$ in Eq.~(\ref{eq:ogt}) 
are one-particle transfer operators, expressed as
\begin{align}
\label{eq:annihilation1}
&\hat P_{j_\nu}
=\zeta_{j_\nu}^{\ast} \tilde a_{{j_\nu}m_{j_{\nu}}}
 + \sum_{j_\nu'} \zeta_{j_\nu j_\nu'}^{\ast} 
s_\nu ({d}_{\nu}^{\+}\times \tilde a_{j_\nu'})^{(j_\nu)}_{m_{j_{\nu}}} \; , \\
\label{eq:annihilation2}
&\hat P_{j_\pi}
=-\theta_{j_{\pi}}^{\ast} s_\pi {a}^{\+}_{j_{\pi}m_{j_{\pi}}}
 - \sum_{j_{\pi}'} \theta_{j_{\pi}j_{\pi}'}^{\ast} 
(\tilde d_{\pi}\times{a}_{j_{\pi}'}^{\+})^{(j_{\pi})}_{m_{j_{\pi}}} \; ,
\end{align}
for the $\btm$ decay of the Zr isotopes. 
The operators in 
Eqs.~(\ref{eq:annihilation1}) and (\ref{eq:annihilation2}), 
respectively, increase 
and decrease the number of like-hole nucleons, 
and both of them decrease the valence nucleon number by 1. 
The coefficients $\zeta_{j}$, $\zeta_{jj'}$, 
$\theta_{j}$, and $\theta_{jj'}$ 
in Eqs.~(\ref{eq:annihilation1}) 
and (\ref{eq:annihilation2}) 
are calculated within the generalized seniority scheme 
\cite{dellagiacoma1988phdthesis}, and depend on 
the $u_j$ and $v_j$ amplitudes. 
Their explicit forms are found in 
Ref.~\cite{nomura2024beta}.

The $ft$ values in seconds are obtained via the 
calculated GT matrix element between the initial 
$I_i$ state of the parent nucleus and the final $I_f$ state 
of the daughter nucleus, 
$\mgt = \braket{I_f \|\hat T^{\rm GT}\|I_i}$, i.e., 
\begin{eqnarray}
 ft = \frac{6163}{\left(\frac{\ga}{\gv}\right)^2 |\mgt|^2} \; ,
\end{eqnarray}
with $\ga=1.27$ and $\gv=1$ being the 
axial-vector and vector coupling constants, 
respectively.

\subsection{Summary of the model parameters}

The parameters for the IBM-2 Hamiltonian 
involved in the present study 
are $\epsilon_d$, $\kappa$, $\kappa_\nu$, 
$\kappa_\pi$, $\chi_\nu$, $\chi_\pi$, 
and $\kappa'$. 
To reduce the number of parameters, 
as in the previous study \cite{nomura2024beta}, 
a simplification is made on the strength parameters for 
the different quadrupole-quadrupole boson interactions, 
so that $\kappa_{\nu}=\kappa_{\pi}=\kappa/2$ for 
the even-even $^{96-102}$Zr nuclei 
and $\kappa_{\nu}=\kappa_{\pi}=0$ for the 
even-even $^{96-102}$Mo nuclei, which are 
taken as the boson cores for the 
odd-odd $^{96-102}$Nb nuclei.  
In addition, in Ref.~\cite{nomura2024beta} 
the $\hat L\cdot \hat L$ term was included 
in the IBM-2 Hamiltonian 
for the even-even Zr nuclei, but was not 
for the even-even Mo nuclei. 
In a number of microscopic 
\cite{OAI,mizusaki1996,nomura2008} and 
phenomenological \cite{IBM} 
IBM-2 calculations carried out to date, however, 
it has been shown that a simplified form of 
the Hamiltonian consisting of the $\hat n_d$, 
and $\hat Q_\nu \cdot \hat Q_\pi$ terms 
describes the low-energy quadrupole 
collective states of most of the medium-heavy and 
heavy nuclei. 
Thus, in the following we regard the parameters 
$\epsilon_d$, $\kappa$, $\chi_\nu$, and $\chi_\pi$ 
the most relevant parameters among the IBM-2 
Hamiltonian (\ref{eq:bham}), and investigate 
dependencies of the results on these parameters, 
concerning the bosonic interactions. 
Tables~\ref{tab:para-zr} and \ref{tab:para-mo} 
summarize, respectively, the IBM-2 parameters 
for the even-even $^{96-102}$Zr and 
odd-odd $^{96-102}$Nb nuclei. 
They were determined microscopically 
by mapping the RHB PES 
onto the corresponding IBM-2 one in Ref.~\cite{nomura2024beta}. 
To avoid confusion, 
we express from now on those IBM-2 
parameters used for the parent Zr nuclei by 
putting a subscript ``$i$'', representing the initial 
state, and those for the daughter Nb nuclei 
with a subscript ``$f$'', so as to represent the 
final state.

Six parameters in the IBFFM-2 Hamiltonian, 
$\Gamma_\nu$, $\Lambda_\nu$, $A_\nu$, 
$\Gamma_\pi$, $\Lambda_\pi$, and $A_\pi$, 
which are the coefficients of the boson-fermion 
interactions, are here treated as free parameters, 
and the variations of the results with these 
parameters are investigated. 
The quasiparticle energies 
$\tilde\epsilon_{\jr}$ in $\hf$, and occupation probabilities 
$v^2_{\jr}$, which appear in $\hbf$ and GT operator, 
are kept constant so as to be the same values 
as those used in the previous study of Ref.~\cite{nomura2024beta}. 
The parameters $\vd$, and $\vt$ in 
the residual interaction $\hff$ are 
additional variable parameters. 
Those $\vd$ and $\vt$ values employed in 
Ref.~\cite{nomura2024beta} are summarized in 
Table~\ref{tab:para-nb}.

%-----------------------------------------------------------
%  
% IBM-2 parameters for Zr
%   
%-----------------------------------------------------------
\begin{table}
\caption{\label{tab:para-zr}
The IBM-2 parameters for the parent nuclei 
$^{96-102}$Zr used in Ref.~\cite{nomura2024beta}, 
which have been obtained from mapping the RHB SCMF 
onto the IBM-2 deformation energy surfaces. 
Note that the like-boson quadrupole-quadrupole interaction 
strengths are assumed to take the values 
$\kappa_{\nu,i}=\kappa_{\pi,i} = \kappa_{i}/2$. 
}
 \begin{center}
 \begin{ruledtabular}
  \begin{tabular}{lccccc}
Nucleus & $\epsilon_{d,i}$ (MeV) & $\kappa_{i}$ (MeV)
& $\chi_{\nu,i}$ & $\chi_{\pi,i}$ & $\kappa'_{i}$ (MeV) \\
\hline
$^{96}$Zr & $0.34$ & $-0.18$ & $-0.35$ & $0.24$ & $0.051$ \\
$^{98}$Zr & $0.28$ & $-0.15$ & $-0.54$ & $0.11$ & $0.032$ \\
$^{100}$Zr & $0.036$ & $-0.094$ & $-0.45$ & $0.20$ & $0.0020$ \\
$^{102}$Zr & $0.081$ & $-0.081$ & $-0.52$ & $0.49$ & $0.0037$ \\
 \end{tabular}
 \end{ruledtabular}
 \end{center}
\end{table}

%-----------------------------------------------------------
%  
% IBM-2 parameters for Nb (Mo)
%   
%-----------------------------------------------------------
\begin{table}
\caption{\label{tab:para-mo}
The boson-core parameters for the odd-odd daughter 
nuclei $^{96-102}$Nb used in Ref.~\cite{nomura2024beta}. 
Note that the strength parameters 
$\kappa_{\pi,i}=\kappa_{\pi,f} = 0$ MeV and 
$\kappa_{f}'=0$ MeV. 
}
 \begin{center}
 \begin{ruledtabular}
  \begin{tabular}{lcccc}
Nucleus & $\epsilon_{d,f}$ (MeV) & $\kappa_f$ (MeV) & $\chi_{\nu,f}$ & $\chi_{\pi,f}$ \\
\hline
$^{96}$Nb & $0.69$ & $-0.44$ & $-0.65$ & $0.45$ \\
$^{98}$Nb & $0.95$ & $-0.41$ & $-0.57$ & $0.08$ \\
$^{100}$Nb & $0.58$ & $-0.35$ & $-0.50$ & $0.45$ \\
$^{102}$Nb & $0.52$ & $-0.27$ & $-0.43$ & $0.36$ \\
 \end{tabular}
 \end{ruledtabular}
 \end{center}
\end{table}

%-----------------------------------------------------------
%     
% Parameters for boson-fermion and fermion-fermion int.
%
%-----------------------------------------------------------
\begin{table}
\caption{\label{tab:para-nb}
Adopted strength parameters (in MeV units) 
for the boson-fermion 
interactions, and fermion-fermion interactions 
in the IBFFM-2 Hamiltonian describing 
the odd-odd $^{96-102}$Nb nuclei. 
The fixed value, $\vd=-0.08$ MeV, 
is employed for the 
$\delta$ term. The spin-spin interaction 
strength, $\vsss=0.1$ MeV is used specifically 
for $^{96}$Nb. 
}
 \begin{center}
 \begin{ruledtabular}
  \begin{tabular}{lccccccc}
Nucleus & $\Gamma_{\nu}$ & $\Gamma_{\pi}$ & $\Lambda_\nu$ & $\Lambda_\pi$ & $A_\nu$ & $A_\pi$ & $\vt$ \\
\hline
$^{96}$Nb & $0.30$ & $0.30$ & $0.40$ & $0.90$& $-0.00$ & $-0.50$ & $0$ \\
$^{98}$Nb & $1.50$ & $0.10$ & $0.80$ & $0.00$& $-1.20$ & $-0.00$ & $0.280$ \\
$^{100}$Nb & $1.50$ & $0.30$ & $0.50$ & $0.20$& $-1.40$ & $-0.80$ & $0.500$ \\
$^{102}$Nb & $1.50$ & $0.10$ & $0.90$ & $3.80$& $-0.90$ & $-2.00$ & $0.500$ \\
 \end{tabular}
 \end{ruledtabular}
 \end{center}
\end{table}

%-----------------------------------------------------------
%
%       log(ft) - IBM-2 parameters in e-e Zr
%
%-----------------------------------------------------------
\begin{figure*}[ht]
\begin{center}
\includegraphics[width=\linewidth]{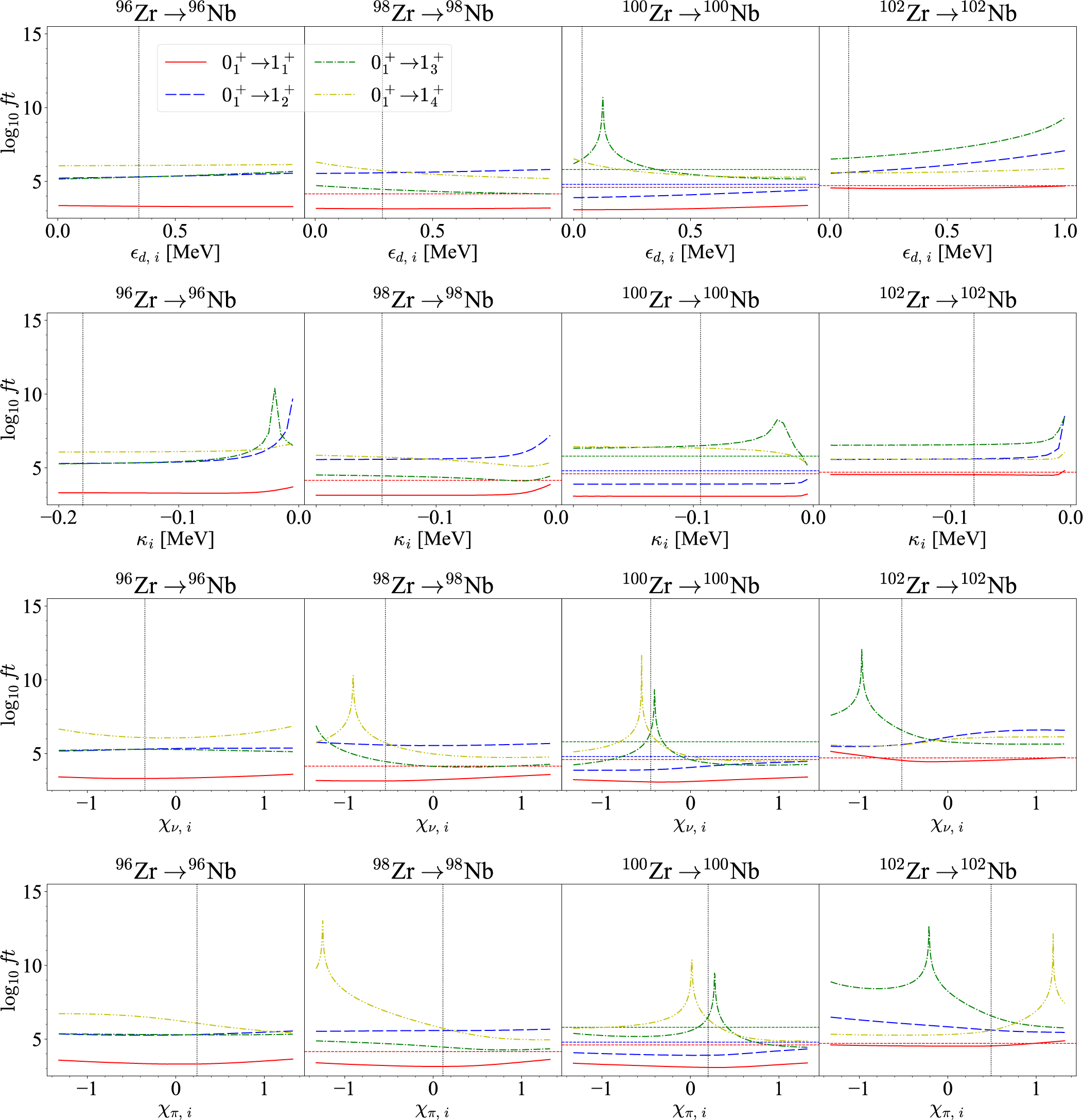}
\caption{Variations of the calculated $\ft$ values 
for the $\btm$ decays of the $0^+_1$ ground states 
of the even-even $^{96-104}$Zr nuclei into 
the lowest four $1^+$ states of the odd-odd 
$^{96-102}$Nb nuclei as 
functions of the IBM-2 parameters for the 
parent Zr nuclei. The vertical dotted line in 
each panel indicates the value of the 
parameter obtained from the RHB-to-IBM 
mapping procedure. Available experimental 
$\ft$ values are also indicated by horizontal 
dashed  lines with the same colors used for the 
corresponding calculated values.}
\label{fig:ft-zr}
\end{center}
\end{figure*}

%-----------------------------------------------------------
%
%       log(ft) - boson-core parameters in o-o Nb
%
%-----------------------------------------------------------
\begin{figure*}[ht]
\begin{center}
\includegraphics[width=\linewidth]{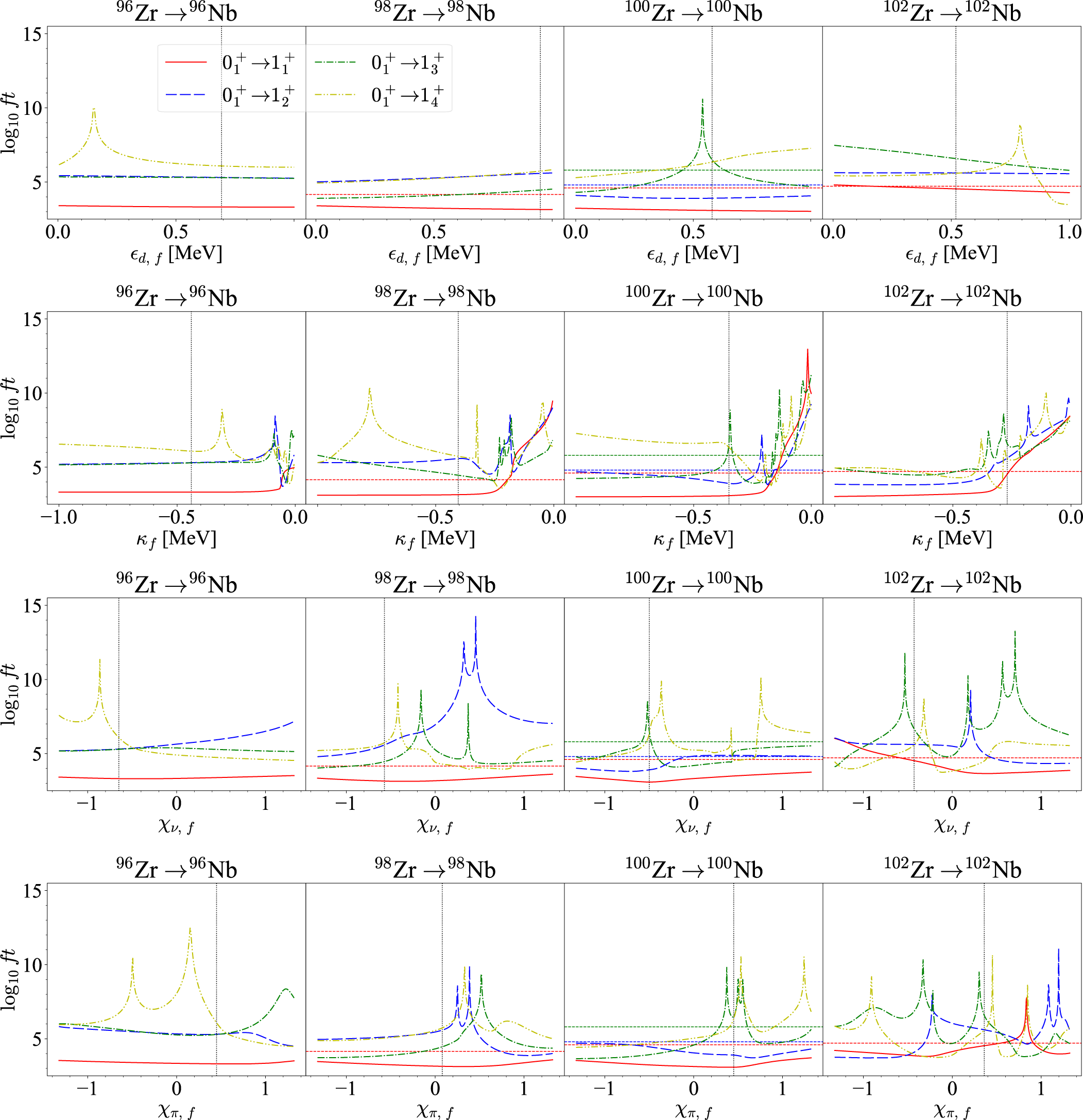}
\caption{Same as Fig.~\ref{fig:ft-zr}, but 
as functions of the boson-core parameters for 
the odd-odd daughter Nb nuclei.}
\label{fig:ft-nb}
\end{center}
\end{figure*}

%-----------------------------------------------------------
%
%       log(ft) - Neutron-boson int.
%
%-----------------------------------------------------------
\begin{figure*}[ht]
\begin{center}
\includegraphics[width=\linewidth]{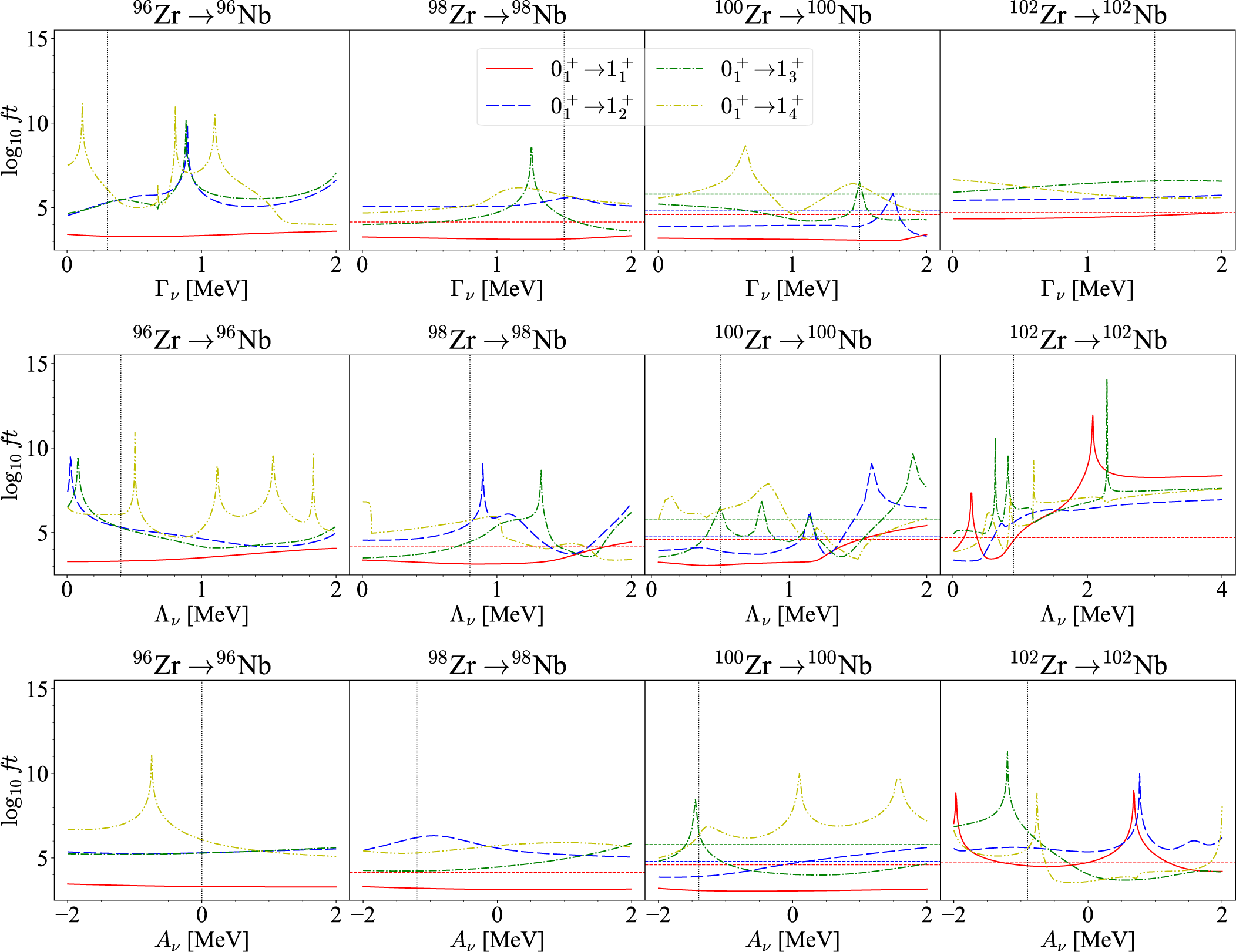}
\caption{Same as Fig.~\ref{fig:ft-zr}, but 
as functions of the interaction strengths between the 
odd neutron and even-even boson core for the odd-odd Nb nuclei.}
\label{fig:ft-bfn}
\end{center}
\end{figure*}

%-----------------------------------------------------------
%
%       log(ft) - Proton-Boson int.
%
%-----------------------------------------------------------
\begin{figure*}[ht]
\begin{center}
\includegraphics[width=\linewidth]{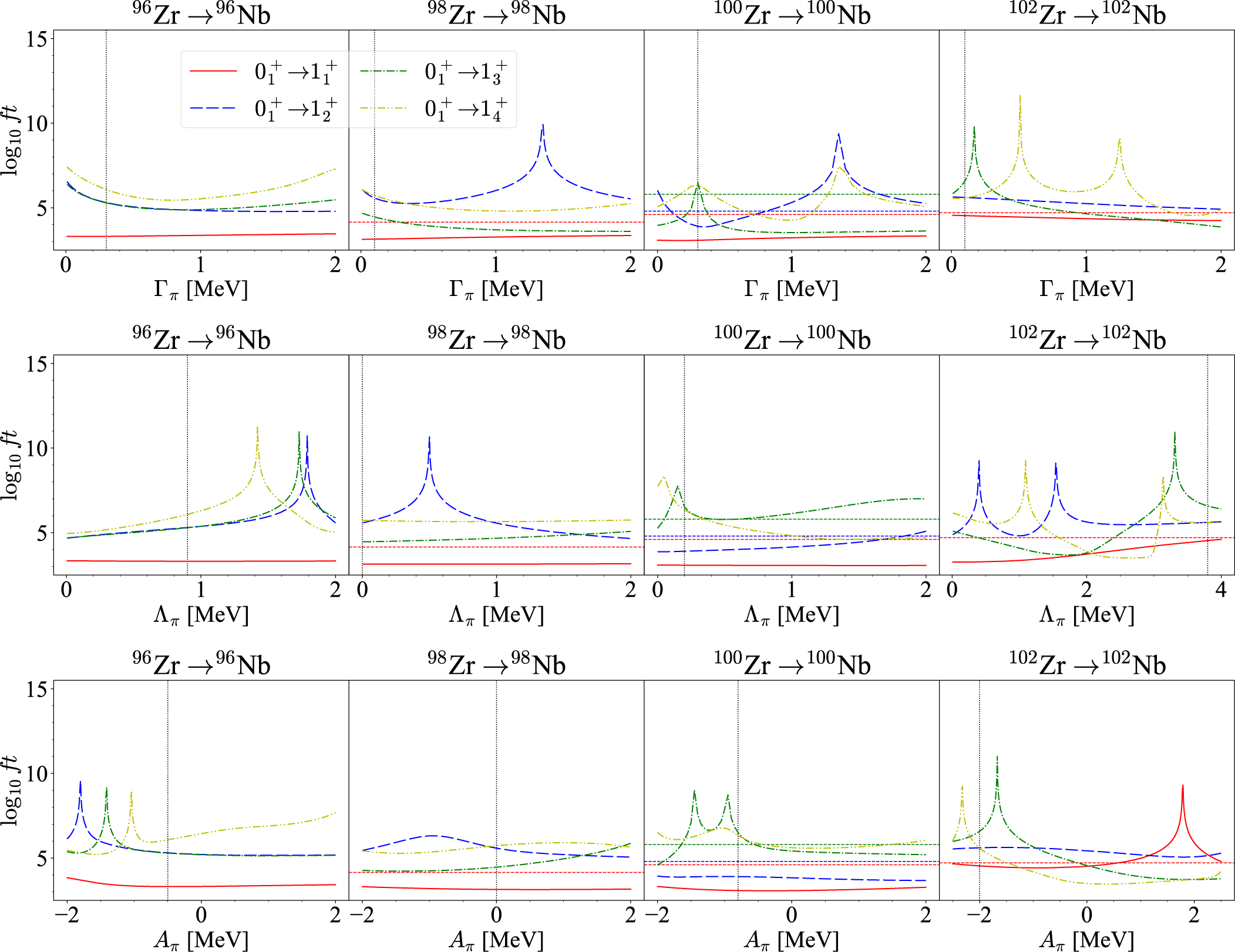}
\caption{Same as Fig.~\ref{fig:ft-zr}, but 
as functions of the interaction strengths between the 
odd proton and even-even boson core for the odd-odd Nb nuclei.}
\label{fig:ft-bfp}
\end{center}
\end{figure*}

%-----------------------------------------------------------
%
%       log(ft) - tensor int.
%
%-----------------------------------------------------------
\begin{figure*}[ht]
\begin{center}
\includegraphics[width=\linewidth]{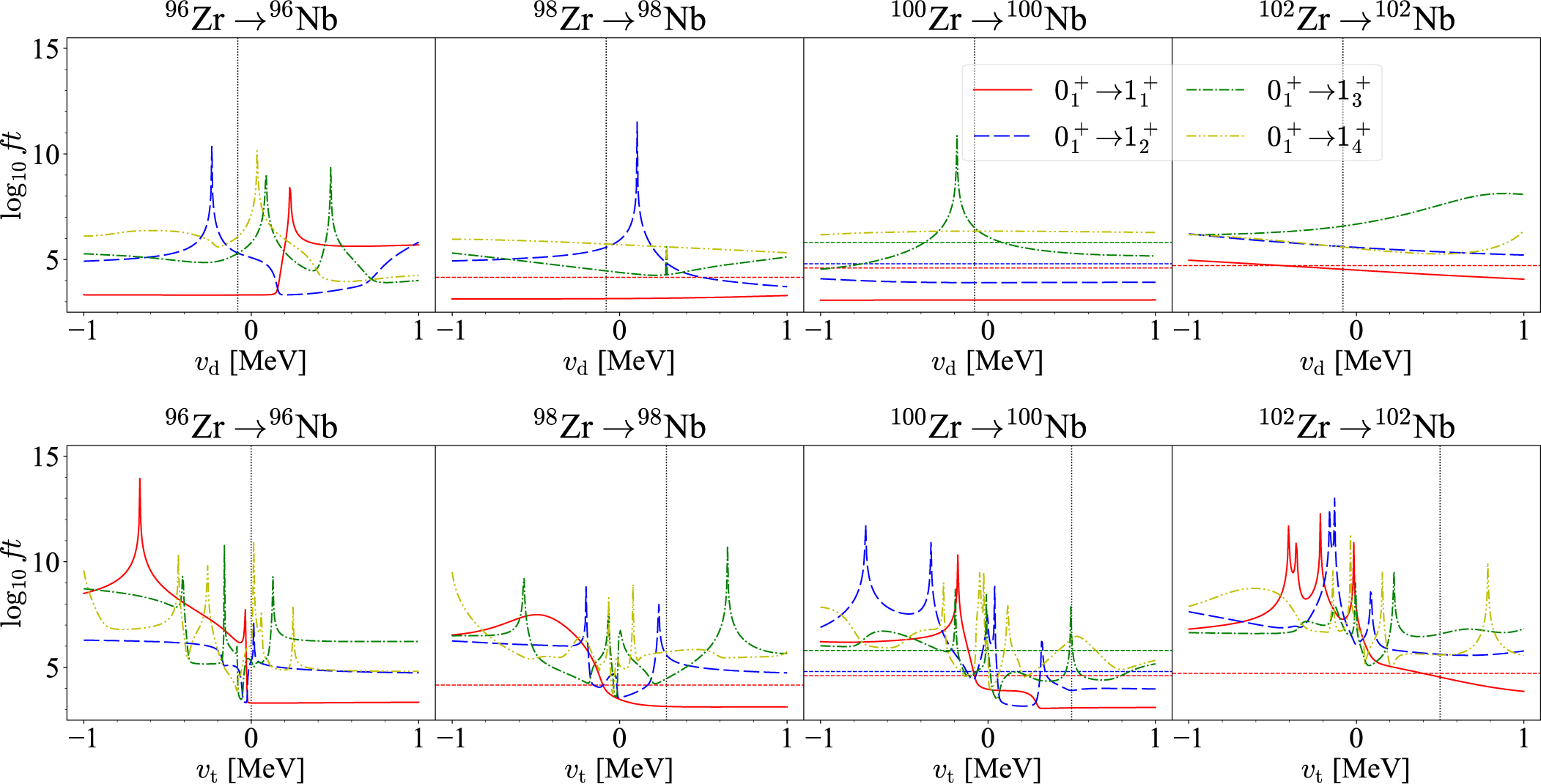}
\caption{Same as Fig.~\ref{fig:ft-zr}, but 
as functions of the strengths parameters for the 
residual neutron-proton interactions considered for the 
odd-odd Nb nuclei.}
\label{fig:ft-vt}
\end{center}
\end{figure*}

%-----------------------------------------------------------
%
%       log(ft) in vt - kappa plane
%
%-----------------------------------------------------------
\begin{figure}[ht]
\begin{center}
\includegraphics[width=\linewidth]{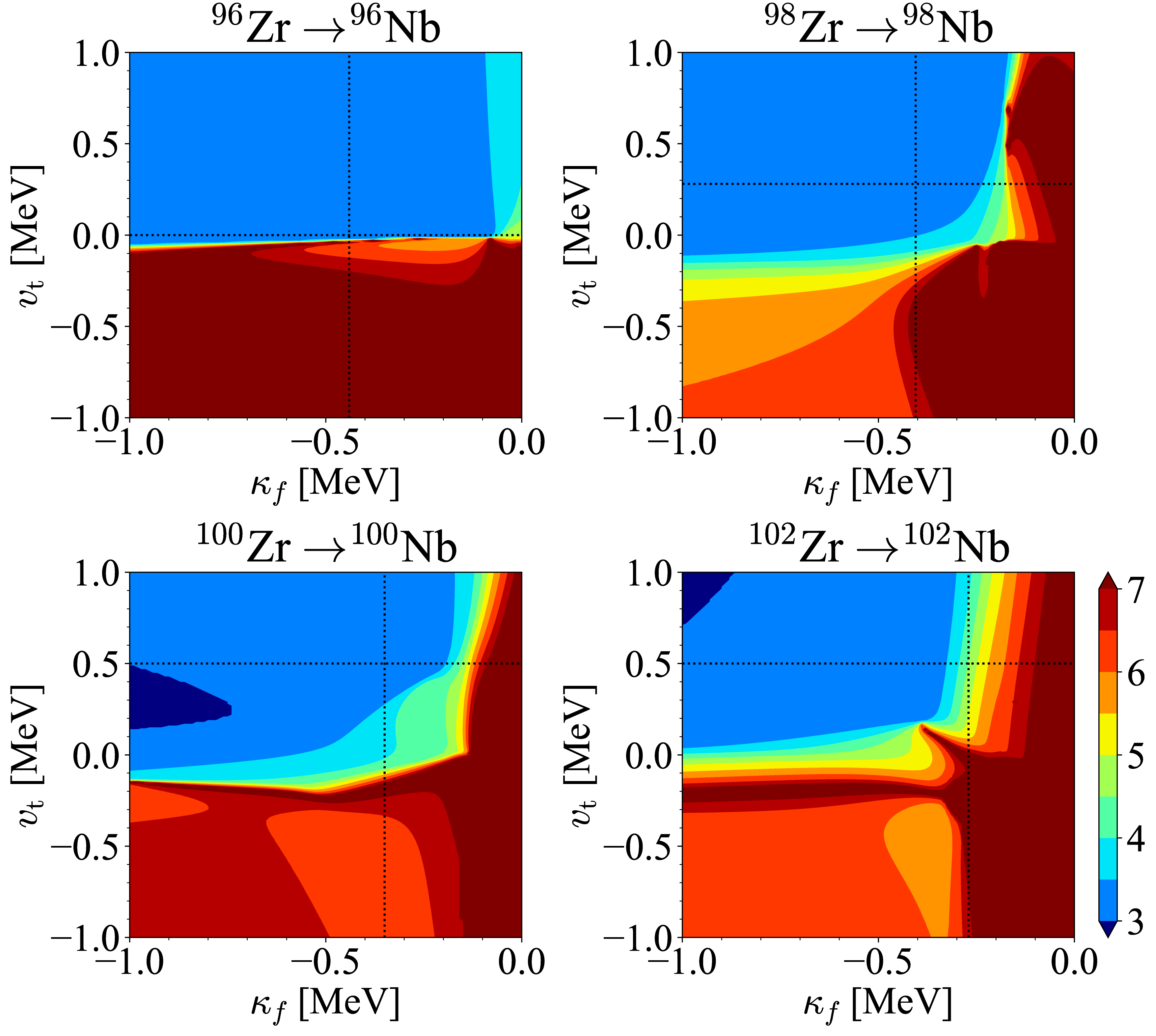}
\caption{Contour plots of the calculated $\btm$-decay 
$\ft(0^+_1 \to 1^+_1)$ values for the even-even $^{96-102}$Zr 
in terms of the parameters $\kappa_f$ and $\vt$ 
used for the odd-odd Nb nuclei. The vertical and 
horizontal dotted lines in each panel 
indicate those $\kappa_f$ and $\vt$ values 
employed in the mapped IBM-2 calculation 
of Ref.~\cite{nomura2024beta}.}
\label{fig:ft-vtkap}
\end{center}
\end{figure}

%-----------------------------------------------------------
%
%       GT transition strength distributions
%
%-----------------------------------------------------------
\begin{figure*}[ht]
\begin{center}
\includegraphics[width=\linewidth]{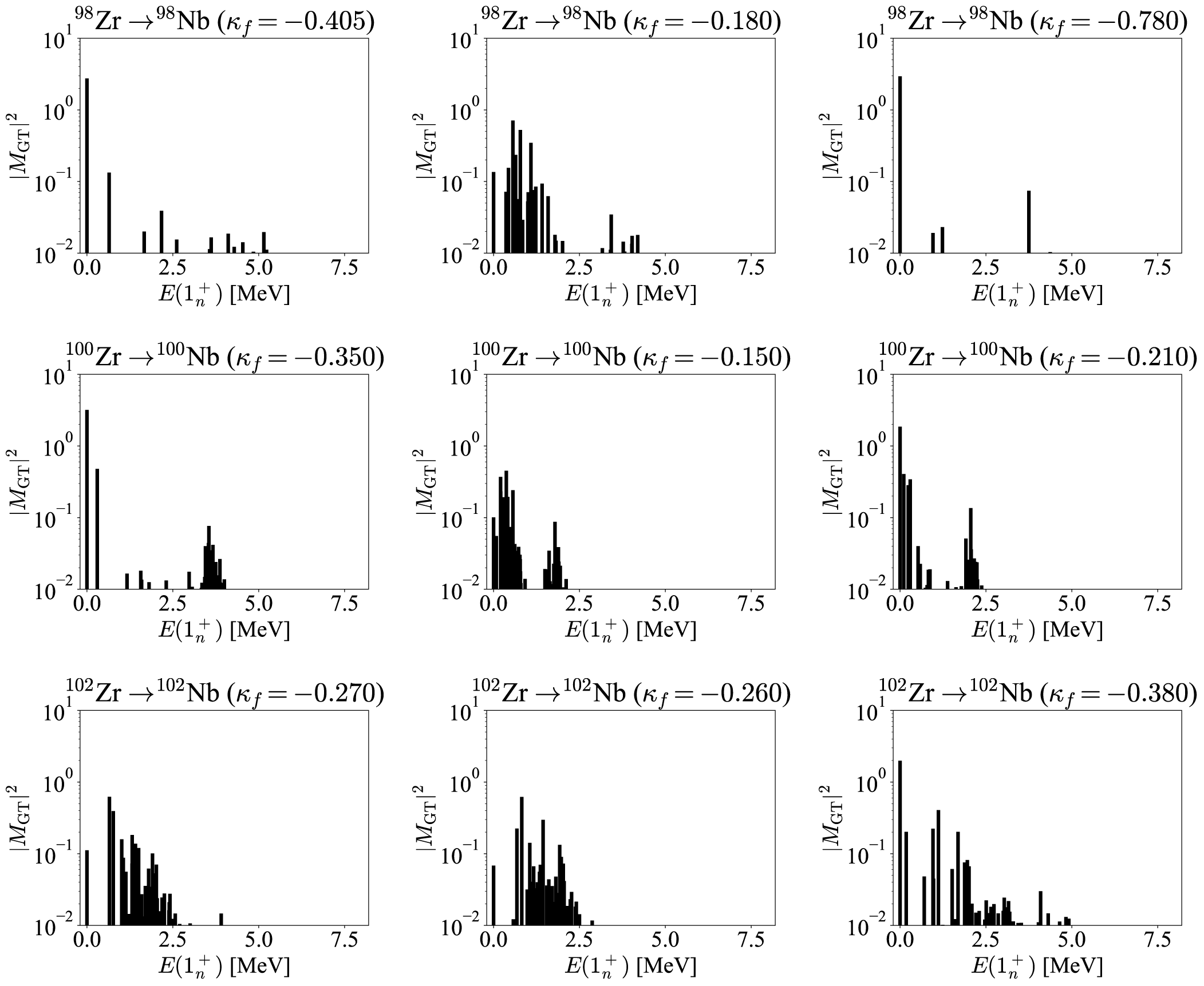}
\caption{Absolute squares of the calculated GT transition 
matrix elements, $|\mgt|^2$, 
for the $^{98,100,102}$Zr$(0^+_1)\to^{98,100,102}$Nb$(1^+)$ 
transitions as functions of the excitation energies 
$E_x(1^+_n)$ of all the $1^+$ states, obtained from 
the IBFFM-2 that employs the $\kappa_f$ values 
(shown in MeV units)
that have been determined from the RHB-to-IBM 
mapping procedure \cite{nomura2024beta} (left column), 
and that are obtained in the present study so as to 
give a reasonable agreement with the experimental 
$\ft(0^+_1 \to 1^+_1)$ value (middle column). 
On the right column shown are the results using those 
$\kappa_f$ value that correspond to a ``spike'' 
pattern in the calculated $\ft$ value. 
See the main text for details. 
}
\label{fig:gt-dist}
\end{center}
\end{figure*}

%-----------------------------------------------------------
%
%       B(GT) running sums
%
%-----------------------------------------------------------
\begin{figure*}[ht]
\begin{center}
\includegraphics[width=\linewidth]{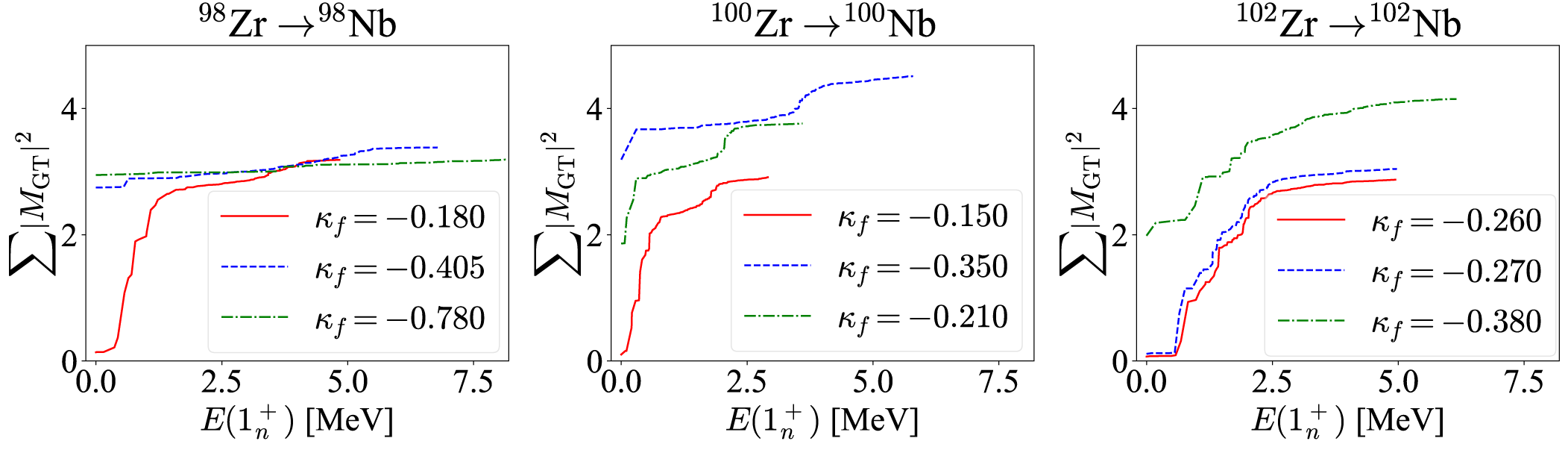}
\caption{Running sums, $|\mgt|^2$, 
as functions of the excitation energies 
$E_x(1^+_n)$ for all the $1^+$ states of the 
odd-odd $^{98,100,102}$Nb nuclei, 
that are obtained from the IBFFM-2 Hamiltonian, 
using the optimal $\kappa_f$ parameter (shown in MeV units)
with which the experimental 
$\ft(0^+_1 \to 1^+_1)$ value is reproduced (solid curves)
and the one determined by the mapping procedure 
\cite{nomura2024beta} (dashed lines). 
The results in the case of the $\kappa_f$ value 
that gives the spike-like pattern in the 
calculated $\ft$ values are also depicted 
as dashed-dotted curves.
}
\label{fig:gt-sum}
\end{center}
\end{figure*}

\section{Impacts on the $\beta$-decay properties\label{sec:ft-para}}

\subsection{$\ft$ values}

Figures~\ref{fig:ft-zr}--\ref{fig:ft-vt} show the calculated 
$\ft$ values for the $\btm$ decay of the $0^+_1$ 
ground state of the even-even $^{96-102}$Zr nuclei into the 
lowest four $1^+$ states of the odd-odd $^{96-102}$Nb 
nuclei as functions of the following model 
parameters: 
the IBM-2 parameters 
in the parent Zr nuclei (Fig.~\ref{fig:ft-zr}), 
the parameters for the even-even boson cores 
(Fig.~\ref{fig:ft-nb}), 
strength parameters for the interactions between 
the odd neutron and boson core (Fig.~\ref{fig:ft-bfn}), 
and between the odd proton and boson core 
(Fig.~\ref{fig:ft-bfp}), and strength parameters 
$\vd$, and $\vt$ for the residual neutron-proton interaction 
(Fig.~\ref{fig:ft-vt}) for the odd-odd daughter nuclei Nb.

One can see from Fig.~\ref{fig:ft-zr} that the 
$\ft(0^+_1 \to 1^+_{1,2})$ values do not depend much 
on all the IBM-2 parameters in the parent Zr nuclei. 
The calculated values $\ft(0^+_1 \to 1^+_{1}) \approx 3.5$ 
do not differ from the ones obtained from the 
mapped IBM-2 parameter (represented by the vertical 
dotted line in each panel), 
and are also much below the 
observed $\ft$ values \cite{data} 
(horizontal dashed lines): 
4.154, 4.6, and 4.71 for 
the $^{98}$Zr$(0^+_1)\to^{98}$Nb$(1^+_1)$, 
$^{100}$Zr$(0^+_1)\to^{100}$Nb$(1^+_1)$, and 
$^{102}$Zr$(0^+_1)\to^{102}$Nb$(1^+_1)$ decays, 
respectively. 
Note that the $\ft$ data are not available for 
the $^{96}$Zr$(0^+_1)\to^{96}$Nb$(1^+_1)$ decay. 
The $\ft(0^+_1 \to 1^+_{3,4})$ values seem to 
be more sensitive to the IBM-2 parameters for 
the parent nuclei.

One should notice that, 
at some specific values of the IBM-2 parameters, 
the calculated $\ft$'s are extremely large, 
as observed, for instance, in 
the $^{96}$Zr$(0^+_1)\to^{96}$Nb$(1^+_3)$ $\btm$ decay 
at $\epsilon_{d,i} \approx 0.1$ MeV. 
The spike-like pattern or discontinuity in 
$\ft$ could be explained 
by the fact that, with a particular combination 
of the parameters,  
matrix elements for different components in the 
GT operator happen to cancel each other 
to such an extent 
that the resultant $\mgt$ matrix element 
nearly vanishes, hence the 
extremely large $\ft$ value is obtained. 
The cancellation of this sort seems to occur 
rather accidentally mainly for the $\btm$ decays to 
higher lying non-yrast $1^+$ states. 
These are, all in all, local behaviors typically 
found at higher $1^+$ excitation 
energies, but do not appear in most of 
the $\ft(0^+_1 \to 1^+_1)$ systematic. 
In addition, as we show in Sec.~\ref{sec:gt}, 
they also do not make any 
sizable contribution to the entire GT strength 
distributions and their 
running sums involving a number of $1^+$ states 
of the odd-odd Nb nuclei. 
Therefore we consider the spike-like patterns of 
$\ft$'s at higher $1^+$ states to be rather unimportant 
in the present analysis, 
especially since we intend to optimize parameters 
by using the $\ft$ data for the lowest-energy, $1^+_1$ state.

In comparison to the behaviors of 
the $\ft$ values as functions 
of the IBM-2 parameters for 
the parent even-even nuclei, one may notice 
in Fig.~\ref{fig:ft-nb} that the calculated 
$\ft$ values are rather sensitive 
to the boson-core Hamiltonian parameters 
for the odd-odd Nb nuclei. 
What is particularly worth noting is the fact that 
the $\ft$ values for the 
$^{98,100,102}$Zr$(0^+_1)\to^{98,100,102}$Nb$(1^+_1)$ decays 
substantially increase with the parameter 
$\kappa_f$, in the range $\kappa_f \gtrsim -0.2$ MeV.
A similar $\ft$ systematic is present for the 
$^{96}$Zr$(0^+_1)\to^{96}$Nb$(1^+_1)$ decay as well 
for the region of even smaller $|\kappa|$ values, 
$\kappa_f \gtrsim -0.1$. 
Similarly, the $^{A}$Zr$(0^+_1)\to^{A}$Nb$(1^+_1)$ decay 
$\ft$ values with $A=98$ and 100 increase with 
the parameter $\kappa_i$ used for the parent Zr nuclei 
(see Fig.~\ref{fig:ft-zr}). 
The increase, however, occurs rather in a narrow 
region of $\kappa_i$, i.e., $\kappa_i \gtrsim -0.01$ MeV, 
which is almost vanishing, and so does not make 
much sense for a realistic calculation. 
One could also extract optimal values for the 
parameter $\kappa_f$, which result in agreement with 
the experimental $\ft(0^+_1 \to 1^+_1)$, when 
available. 
For those $^{A}$Zr$(0^+_1)\to^{A}$Nb$(1^+_1)$ decays 
with $A=98$, 100, and 102, the $\kappa_f$ values of 
approximately $-0.180$, $-0.150$, and $-0.260$ MeV 
give a good description of the corresponding $\ft$ data. 
The above values correspond to the reduction 
in magnitude of $\approx 55$ \% of 
those adopted in the mapped IBM-2 
calculations \cite{nomura2024beta}.
The calculated $\ft$ values do not show 
a significant dependence on the 
parameters $\chi_{\nu,f}$ and $\chi_{\pi,f}$, 
an exception being perhaps 
the $^{102}$Zr$(0^+_1)\to^{102}$Nb$(1^+_1)$ decay.

We notice in Fig.~\ref{fig:ft-bfn} 
that the predicted $\ft(0^+_1 \to 1^+_1)$ values 
as a function of $\Gamma_{\nu}$ 
(the dynamical quadrupole interaction 
strength between the odd neutron and 
even-even boson core) 
are stable for most of the 
nuclei. A stronger parameter dependence of the $ft$ 
values is seen in their evolution with the exchange 
interaction strength $\Lambda_\nu$ for 
the $^{A}$Zr$(0^+_1)\to^{A}$Nb$(1^+_1)$ decay with 
$A=98$, 100, and 102. 
It is worth noting that 
the exchange interaction reflects the fact that 
the bosons are made of pairs of fermions, and 
this type of the interaction has been shown 
to play an important role in reproducing 
low-energy levels of odd nuclei.  
It appears from the systematic shown in 
Fig.~\ref{fig:ft-bfn} that the exchange interaction 
also has an impact on the calculation of 
the $\beta$-decay $\ft$ values for 
$A=98$, 100, and 102, as well as 
the low-lying states. 
The monopole interaction strength $A_\nu$ does not 
seem to have an influence on the $\ft$ calculations. 
It is more or less expected from 
the fact that this interaction only has 
an effect of either compressing or stretching a 
whole energy spectrum, and hence plays a less 
important role than the dynamical and exchange 
interactions. 
An exception is an irregular, spike-like, 
behavior of $\ft(0^+_1 \to 1^+_1)$ 
at $A_\nu \approx -2$ MeV and $\approx 1.75$ MeV
in the case of the $^{102}$Zr$\to^{102}$Nb decay.

It appears from Fig.~\ref{fig:ft-bfp} that 
the $\ft$ values are less sensitive to 
$\Gamma_{\pi}$ 
(the strength parameters for the interaction 
between the odd proton and even-even boson core), 
as the $\ft(0^+_1 \to 1^+_1)$ value 
is almost constant against the variations of the 
$\Gamma_\pi$, $\Lambda_\pi$, and $A_{\pi}$ 
parameters for most of the considered decay processes 
$^{A}$Zr$\to^{A}$Nb with $A=96$, 98, and 100. 
One notices a certain dependence of the 
$^{102}$Zr$(0^+_1)\to^{102}$Nb$(1^+_1)$ $\ft$ value 
on $\Lambda_\pi$, 
as it smoothly increases with 
$\Lambda_\pi \gtrsim 1$ MeV.

In Fig.~\ref{fig:ft-vt}, we observe 
a strong dependence 
of the $^{96}$Zr$(0^+_1)\to^{96}$Nb$(1^+_n)$ ($n=1-4$) 
decay $\ft$ values on the parameter $\vd$, 
a residual neutron-proton 
interaction strength of surface-$\delta$ type, 
seen most spectacularly in the sharp rise 
of the $\ft\,(0^+_1 \to 1^+_1)$ value near 
$\vd= 0.2$ MeV. 
For the $^{98,100}$Zr$(0^+_1)\to^{98,100}$Nb$(1^+_n)$ 
decays, in contrast, the corresponding 
$\ft$ values exhibit a much weaker 
dependence on $\vd$. 
The $^{102}$Zr$(0^+_1)\to^{102}$Nb$(1^+_{1,2})$ decay $\ft$ 
values show a decreasing pattern as functions of $\vd$, 
but the change is smooth and monotonic, 
as compared to the case of the 
$^{96}$Zr$(0^+_1)\to^{96}$Nb$(1^+_{1,2})$ decay $\ft$ values. 
Such a specialty for the 
$^{96}$Zr$(0^+_1)\to^{96}$Nb$(1^+_{n})$ decays 
is perhaps due to the fact that among the odd-odd Nb nuclei 
studied here 
only $^{96}$Nb has the ground state with 
spin and parity $I^\pi=6^+$, 
while the others have $I^{\pi}=1^+$; 
to reproduce the $6^+$ ground state, 
in Ref.~\cite{nomura2024beta} 
a different form of the residual 
neutron-proton interactions was considered for $^{96}$Nb, 
so that $\vt=0$ MeV and a spin-spin interaction 
of the form 
$-{\bm\sigma}_\nu\cdot {\bm\sigma}_\pi/\sqrt{3}$
was specifically included for 
this nucleus (see Ref.~\cite{nomura2024beta} and 
the caption to Table~\ref{tab:para-nb}). 
As is clear from Fig.~\ref{fig:ft-vt}, the 
predicted $\ft$ values, particular for the 
$^{A}$Zr$(0^+_1)\to^{A}$Nb$(1^+_{1,2})$ decays of 
all the masses $A=96-102$, 
consistently exhibit a strong dependence on 
$\vt$, in such a way that 
they rise sharply, as $\vt$ decreases 
for $\vt<0$. 
For the 
$^{98,100}$Zr$(0^+_1)\to^{98,100}$Nb$(1^+_{1})$ decays, 
the value $\vt\approx -0.1 (<0)$ MeV appears to lead to 
a good agreement with the experimental data, 
$\ft\approx 4.154$ and 4.6, respectively. 
Concerning the 
$^{102}$Zr$(0^+_1)\to^{102}$Nb$(1^+_{1})$ decay, 
as it is seen from Fig.~\ref{fig:ft-vt} the 
derived $\vt=0.5$ MeV in the mapped IBM-2 framework 
already gives a reasonable agreement with 
the experimental value of $\ft=4.71$.

So far we have seen that the $\ft(0^+_1 \to 1^+_1)$ 
values for the considered $^A$Zr$\to^A$Nb $\btm$ decays 
with $A=96-102$ all show a particularly strong dependence on the 
parameters $\kappa_f$ and $\vt$.  
We thus analyze the behaviors of the 
$^{A}$Zr$(0^+_1)\to^{A}$Nb$(1^+_{1})$ 
decay $\ft$ values 
when we change $\kappa_f$ and $\vt$ simultaneously. 
The results are presented in Fig.~\ref{fig:ft-vtkap}, 
where the calculated $\ft(0^+_1 \to 1^+_1)$ values are 
depicted in contour plots. 
In the figure, 
the $^{96}$Zr$(0^+_1)\to^{96}$Nb$(1^+_{1})$ decay $\ft$ 
looks almost constant against $\kappa_f$, since only in the 
vicinity of $\kappa_f=0$ MeV is a slight increase 
observed for $\vt>0$ and some weak dependence on $\kappa_f$ 
within the range $-0.6 \lesssim \kappa_f \lesssim -0.1$ MeV. 
When it is seen as a function of $\vt$, there appears 
a sharp decrease when it changes sign 
from $\vt<0$ to $\vt>0$. 
The $\btm$ decays of the $^{98,100,102}$Zr nuclei 
are more interesting here, since there are 
experimental $\ft\,(0^+_1 \to 1^+_1)$ data available. 
As compared to the $^{96}$Zr$(0^+_1)\to^{96}$Nb$(1^+_{1})$ 
decay $\ft$ value, those for the 
$^{98,100,102}$Zr$(0^+_1)\to^{98,100,102}$Nb$(1^+_{1})$ decays 
show a notable dependence 
on both the $\kappa_f$ and $\vt$ parameters. 
The predicted $\ft(0^+_1 \to 1^+_1)$ values 
obtained from the RHB mapped-IBM-2 calculation 
\cite{nomura2024beta}, are shown as the 
crossing point of the vertical and horizontal 
dotted lines in each panel of Fig.~\ref{fig:ft-vtkap}, 
representing, respectively, 
those $\kappa_f$ and $\vt$ parameters 
employed in that calculation. 
We notice that they are rather far from 
the observed $\ft$ values, 
4.154, 4.6, and 4.71 for the 
$^{98,100,102}$Zr$(0^+_1)\to^{98,100,102}$Nb$(1^+_{1})$ decays, 
respectively. 
Based on the $(\kappa_f,\vt)$ surfaces 
in Fig.~\ref{fig:ft-vtkap}, 
we can extract optimal sets of the 
$\kappa_f$ and $\vt$ values 
to improve description of the $\ft$ data 
available for the $^{98,100,102}$Zr $\btm$ decays. 
For the $^{A}$Zr$(0^+_1)\to^{A}$Nb$(1^+_{1})$ decays 
with $A=98$, 100, and 102, one could choose 
$\kappa_f=-0.180$, $-0.150$, and $-0.260$ MeV, 
respectively. Given these $\kappa_f$ values, 
the $\ft(0^+_1 \to 1^+_1)$ appears to 
be rather insensitive to $\vt$ in 
the $(\kappa_f,\vt)$ surface, so it would 
be just enough to employ the same $\vt$ values 
as those used in the mapped IBM-2 calculations, 
i.e., $\vt=0.28$, 0.5, and 0.5 MeV for 
the $^{98}$Zr, $^{100}$Zr, and $^{102}$Zr 
$\btm$ decays, respectively (cf. Table~\ref{tab:para-nb}).

\subsection{GT strength distributions\label{sec:gt}}

Figure~\ref{fig:gt-dist} displays absolute squares of 
the calculated GT 
transition matrix elements, $|\mgt(0^+_1 \to 1^+_n)|^2$, 
for the $\btm$ decays of the even-even $^{96-102}$Zr into 
odd-odd $^{98-102}$Nb nuclei 
as functions of the excitation energies $E_x(1^+_n)$ of all 
the $1^+$ states of the odd-odd daughter (Nb) nuclei, 
obtained from the IBFFM-2 Hamiltonian 
in the considered model space. 
The calculated results employing different values of the 
strength parameter $\kappa_f$ for the odd-odd Nb nuclei 
are compared: the derived value from the 
RHB-to-IBM mapping in Ref.~\cite{nomura2024beta} (left column), 
and the optimal value extracted in the present work, 
that gives an agreement with the 
measured $\ft(0^+_1 \to 1^+_1)$ (middle column). 
Also included on the right column of 
Fig.~\ref{fig:gt-dist} are 
the results obtained with 
the $\kappa_f$ values that give rise the 
``spike'' pattern we observe in Fig.~\ref{fig:ft-nb} 
in the calculated $\ft$ values 
for non-yrast $1^+$ states. 
Specifically, 
at the values $\kappa_f=-0.780$, $-0.210$, and $-0.380$ MeV, 
the calculated $\ft(0^+_1 \to 1^+_4)$, 
$\ft(0^+_1 \to 1^+_3)$, and $\ft(0^+_1 \to 1^+_4)$ 
values for the $\btm$ decays of 
$^{98,100,102}$Zr, respectively, 
are extremely large (see Fig.~\ref{fig:ft-nb}).

Note that the mapped IBM-2 results for 
the $\mgt$ distributions were already presented 
in Fig.~9 of Ref.~\cite{nomura2024beta}. 
In that reference, 
the maximal number of iterations in the 
numerical diagonalization of the IBFFM-2 Hamiltonian 
using the Lanczos method 
were set to be 20000, 200000, and 200, 
for $^{98}$Nb, $^{100}$Nb, and $^{102}$Nb 
nuclei, respectively. 
In the present study, 
the iterations are carried out 200000 times 
for all the Nb nuclei considered to achieve a better 
convergence. 
There is no noticeable difference between 
the previous result using the smaller number of 
the iterations, and the present one in the systematic 
of the GT transition distributions. 
An exception is the behaviors of 
the GT strengths for the 
high $1^+$ excitation energies, which, however, 
only make negligible contributions 
to the gross feature of the $\mgt$ 
distributions and their running sums.

It is seen from Fig.~\ref{fig:gt-dist} that, 
in general, the GT transitions to low-lying $1^+$ states 
below $E_x(1^+)\approx 1$ MeV 
make the dominant contribution to the $\mgt$ strengths. 
The calculated $\mgt$ values also seem to 
be sensitive to the $\kappa_f$ value, in such a way that, 
as $|\kappa_f|$ decreases, 
the GT strengths look more 
densely populated in the lower energy 
region of $E_x(1^+)$, and are also more fragmented. 
For the $^{98}$Zr$(0^+_1)\to^{98}$Nb$(1^+)$ 
decay, the corresponding $\mgt$ distribution pattern with 
$\kappa_f=-0.780$ MeV more or less resembles 
the one obtained with $\kappa_f=-0.405$ MeV, 
since in both cases 
the total $\mgt$ sum would be mainly 
accounted for by the GT transition to the 
lowest energy $1^+$ state. 
The overall patterns of the GT strengths 
for the $^{100}$Zr$(0^+_1)\to^{100}$Nb$(1^+)$ 
decay are qualitatively similar between the calculations 
employing the three different values of 
$\kappa_f$, in that 
mainly two peaks appear below $E_x(1^+)\approx 1$ MeV, and near 
or above $E_x(1^+)\approx 1$ MeV. 
Regarding the $^{102}$Zr$(0^+_1)\to^{102}$Nb$(1^+)$ $\btm$ 
decay, the systematic behaviors of the 
$\mgt$ distributions with different $\kappa_f$ values 
are similar to each other, as well.

In Fig.~\ref{fig:gt-sum} we show running sums of the 
GT transition strengths $\sum_n |\mgt(0^+_1 \to 1^+_n)|^2$ 
for the $^{98,100,102}$Zr$(0^+_1)\to^{98,100,102}$Nb$(1^+)$ 
$\btm$ decays 
with respect to $E_x(1^+)$, 
calculated with three different values of the parameter $\kappa_f$. 
For all the transitions shown in the figure, 
one can observe a systematic trend that, 
for larger $|\kappa_f|$ the GT running sum is 
mostly accounted for by the transitions to 
lowest-lying $1^+$ states. This is also confirmed 
from the $|\mgt|^2$ systematic in Fig.~\ref{fig:gt-dist}. 
For the $^{98}$Zr$(0^+_1)\to^{98}$Nb$(1^+)$ decay, 
$\sum_n|\mgt|^2$ with $\kappa_f=-0.405$ and $-0.780$ MeV 
are shown to be converged already 
near the lowest energy, $1^+_1$ state. 
In the case of $\kappa_f=-0.180$ MeV, 
on the other hand,  
contributions from the lowest $1^+$ states 
are negligible. 
The sum rises at $E_x(1^+) \approx 1$ MeV, and 
converges to $\sum_n|\mgt|^2 \approx 3$ at $E_x(1^+)\approx 4$ MeV. 
The calculated GT sums with the three 
different $\kappa_f$'s seem to converge consistently 
to $\sum_n|\mgt|^2 \approx 3$, meaning that the 
GT sum does not depend much on the 
parameter $\kappa_f$ for the 
$^{98}$Zr$(0^+_1)\to^{98}$Nb$(1^+)$ decay.

This finding appears to be rather at variance 
with the results for the 
$^{100,102}$Zr$(0^+_1)\to^{100,102}$Nb$(1^+)$ decays. 
In these cases, 
as we decrease $|\kappa_f|$, the running sum 
$\sum_n|\mgt|^2$ converges at low $E_x(1^+)$ energies, 
and the converged value becomes  
smaller, e.g., 
$\sum_n|\mgt|^2 \approx 4.5$, 3.7, and 2.8 for 
the $^{100}$Zr$(0^+_1)\to^{100}$Nb$(1^+)$ decay 
with $\kappa_f=-0.350$, $-0.210$, and $-0.150$ MeV, 
respectively. 
The variation of $\kappa_f$ is, therefore, shown to 
affect much the $\sum_n|\mgt|^2$ sums, as regards 
the $^{100,102}$Zr$(0^+_1)\to^{100,102}$Nb$(1^+)$ decays.

%-----------------------------------------------------------
%
%       Nb energies in vt - kappa plane
%
%-----------------------------------------------------------
\begin{figure*}[ht]
\begin{center}
\includegraphics[width=.8\linewidth]{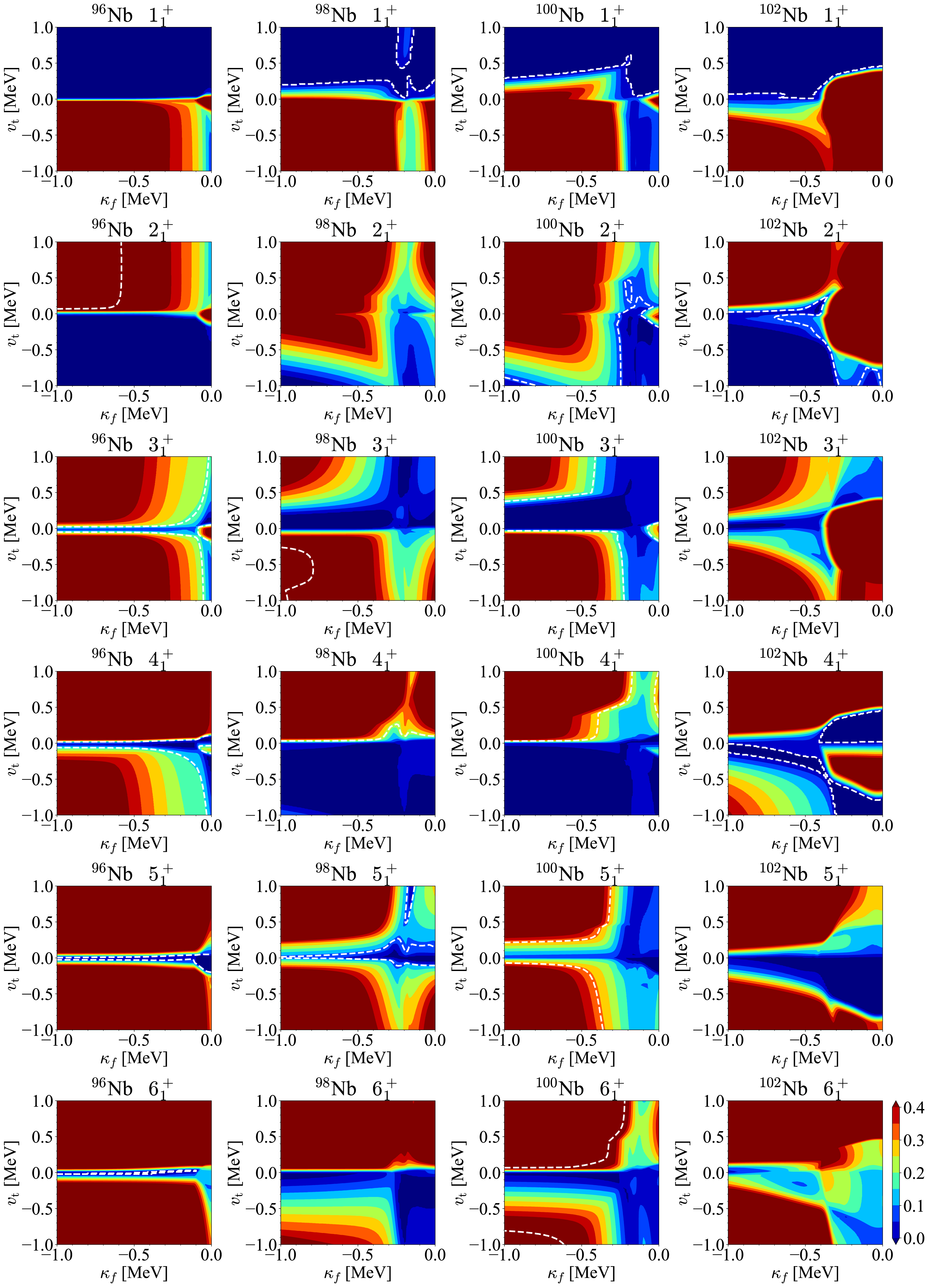}
\caption{Contour plots of the calculated excitation 
energies for the yrast $1^+$, $2^+$, $3^+$, $4^+$, 
$5^+$, and $6^+$ states of the daughter nuclei Nb 
shown as functions of the parameters $\kappa_f$ and 
$\vt$ used for the odd-odd Nb nuclei. 
The dashed lines in each panel connect values 
corresponding to the experimental excitation energies.}
\label{fig:e-vtkap}
\end{center}
\end{figure*}

%-----------------------------------------------------------
%
%       E_x(1+) and log(ft) trajectory
%
%-----------------------------------------------------------
\begin{figure*}[ht]
\begin{center}
\includegraphics[width=\linewidth]{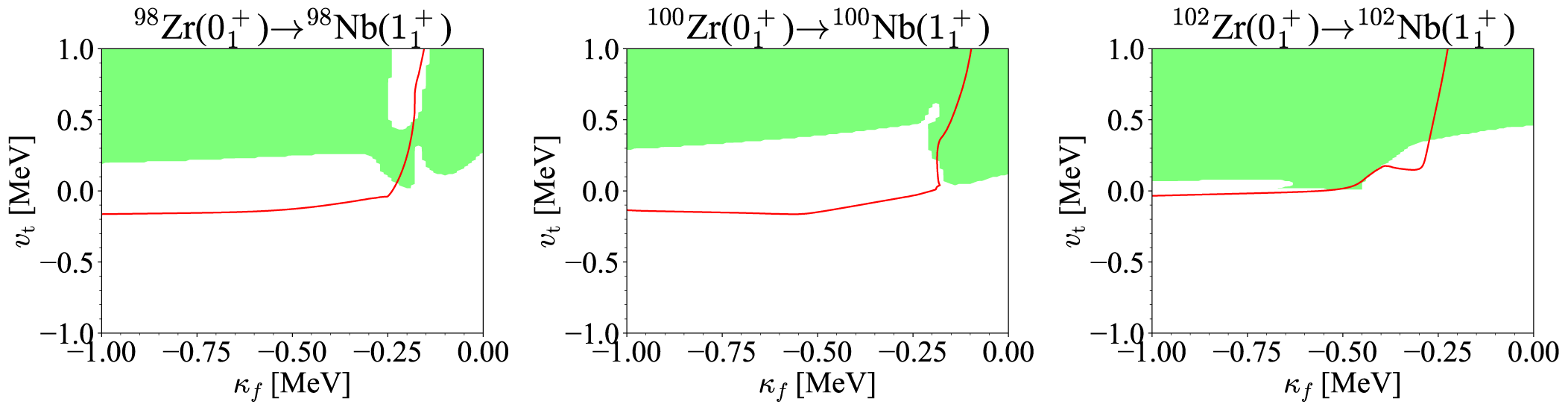}
\caption{Regions of those $\kappa_f$ and $\vt$ parameters 
(shaded areas) that give the $1^+_1$ state 
to be the ground state for $^{98,100,102}$Nb. 
The solid curves connects the $(\kappa_f,\vt)$ values 
with which the experimental $\ft(0^+_1 \to 1^+_1)$ 
values for the Zr $\to$ Nb $\btm$ decay \cite{data} are reproduced.}
\label{fig:e-ft}
\end{center}
\end{figure*}

%-----------------------------------------------------------
%
%       Comparison of the Nb energy level
%
%-----------------------------------------------------------
\begin{figure*}[ht]
\begin{center}
\includegraphics[width=\linewidth]{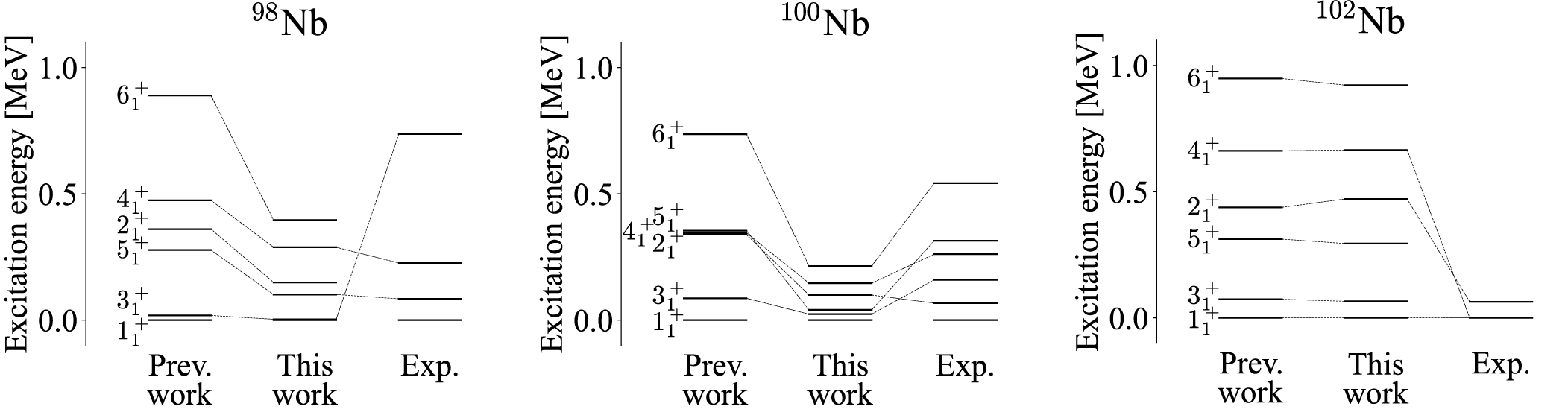}
\caption{Comparison of the low-energy spectra for the 
odd-odd $^{98,100,102}$Nb nuclei calculated with the parameters that 
give a reasonable agreement with the experimental 
$\ft(0^+_1 \to 1^+_1)$ data to those obtained 
in Ref.~\cite{nomura2024beta}, and to the experimental 
data \cite{data}.}
\label{fig:level-nb}
\end{center}
\end{figure*}

%-----------------------------------------------------------
%
%       Comparison of the Mo energy level
%
%-----------------------------------------------------------
\begin{figure*}[ht]
\begin{center}
\includegraphics[width=\linewidth]{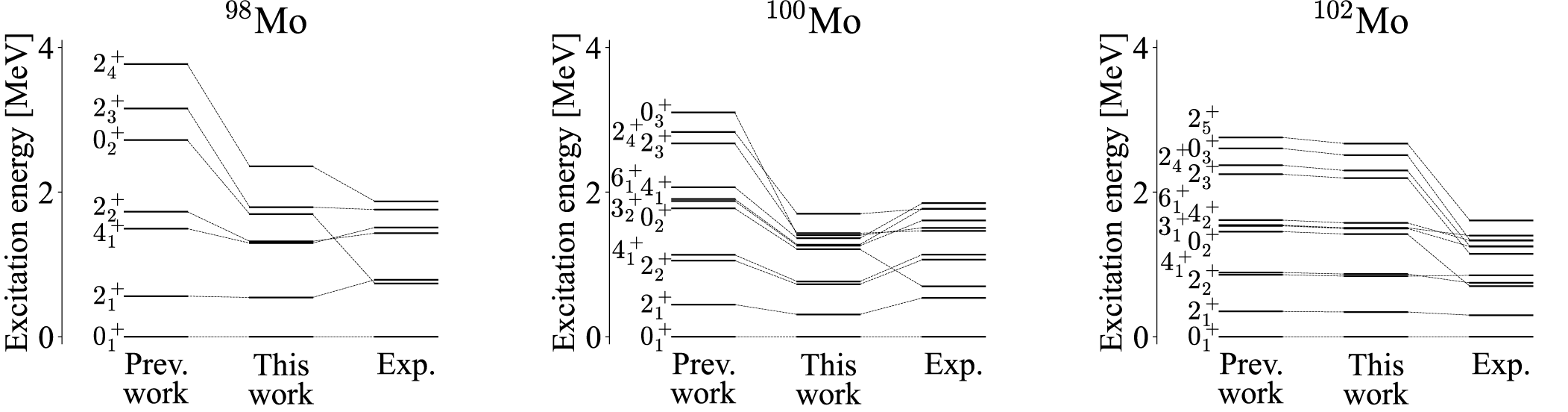}
\caption{Same as Fig.~\ref{fig:level-nb}, but for the 
energy spectra for the even-even $^{98,100,102}$Mo nuclei.}
\label{fig:level-mo}
\end{center}
\end{figure*}

\section{Impacts on the energy spectra\label{sec:energy-para}}

We now turn to discuss the parameter dependence of  
the results on the low-lying states of 
each nucleus involved, 
specifically focusing on the energy 
levels of the odd-odd Nb nuclei. 
Given the fact that among all the model parameters 
involved in the $\ft$ calculations, the strength 
parameters $\kappa_f$ and $\vt$ most affect 
the $\ft(0^+_1 \to 1^+_1)$ predictions 
(cf. Figs.~\ref{fig:ft-nb} and \ref{fig:ft-vt}), 
we analyze in particular the behaviors of low-spin yrast 
states of the odd-odd Nb nuclei when these 
parameters are varied simultaneously. 
We show in Fig.~\ref{fig:e-vtkap} contour plots 
of the calculated $1^+_1$, $2^+_1$, $3^+_1$, $4^+_1$, 
$5^+_1$, and $6^+_1$ states of the odd-odd 
$^{96-102}$Nb nuclei in terms of the parameters 
$\kappa_f$ and $\vt$. 
In each plot, those $(\kappa_f,\vt)$ values giving the 
excitation energies that agree with the available 
experimental data are connected by the dashed lines.

As mentioned earlier, 
the measured ground state of $^{96}$Nb 
has the spin and parity $I^{\pi}=6^+$ \cite{data}. 
One sees in Fig.~\ref{fig:e-vtkap} that 
the $5^+_1$ and $6^+_1$ excitation energies 
of $^{96}$Nb are sensitive to $\kappa_f$ 
within the range $\kappa_f \gtrsim -0.2$, 
and to $\vt$ with its value being near $\vt=0$ MeV. 
Optimal $(\kappa_f,\vt)$ values that are to 
reproduce the correct $6^+$ ground-state spin 
should also be extracted somewhere from 
these parameter ranges.

As is shown in Fig.~\ref{fig:ft-vtkap}, 
overall behaviors of the excitation 
energies within the $(\kappa_f,\vt)$ plane look similar 
between $^{98}$Nb and $^{100}$Nb, which are, however, 
rather different from the one for the $^{102}$Nb nucleus. 
The calculated energies for the heavier Nb nuclei, 
$^{98-102}$Nb, appear to depend on these parameters more 
strongly than in the case of $^{96}$Nb, e.g., 
for the $5^+_1$ and $6^+_1$ states. 
The ground state for $^{98,100,102}$Nb 
is experimentally suggested to be the $1^+$ state. 
As one sees from the behaviors of the dashed curves in 
Fig.~\ref{fig:e-vtkap}, $\vt$ should take a positive value 
so that the measured ground-state 
spin should be reproduced correctly. 
Given a positive $\vt$ value, then typical $\kappa_f$ values 
that reproduce the excitation energies of the 
states other than $1^+_1$ may be, perhaps, 
within the range $-0.3 \lesssim \kappa_f \lesssim -0.2$ MeV, 
in which the energies significantly depend on $\kappa_f$. 
It is, however, not very obvious to find a best set 
of the $(\kappa_f,\vt)$ values that 
reproduces all the excitation energies, as well as 
the observed $\ft(0^+_1 \to 1^+_1)$ values 
satisfactorily. 
Other model parameters, e.g., those for the spin-spin-$\delta$ 
and spin-spin interactions for the residual neutron-proton 
correlations, may need to be taken into account as 
other adjustable parameters. 
Such an analysis would invoke further complications, 
and is beyond the scope of the present study.

To shed light upon the search for optimal sets 
of the $(\kappa_f,\vt)$ parameters, 
we now consider as constraints the $\ft(0^+_1 \to 1^+_1)$ data, 
and the ground-state spin of $1^+$ for the 
odd-odd $^{98,100,102}$Nb nuclei. 
Figure~\ref{fig:e-ft} indicates regions of those values 
of the parameters $\kappa_f$ and $\vt$, shown as shaded areas, 
that give the $1^+_1$ state to be the ground state 
of the odd-odd final nuclei for the 
$0^+_1 \to 1^+_1$ GT transitions of the $^{98}$Zr, 
$^{100}$Zr, and $^{102}$Zr nuclei. 
Also indicated in Fig.~\ref{fig:e-ft} 
are those $(\kappa_f,\vt)$ values, 
which are connected by solid curves, 
that reproduce the observed $\ft$ value for each 
of the $^{A}$Zr$(0^+_1)\to^{A}$Nb$(1^+_1)$ $\btm$ decays. 
Optimal sets of the $(\kappa_f,\vt)$ values 
would be those at which the 
$\ft$ curve passes through the shaded 
area in the figure. 
They could be 
$0.1 \lesssim \vt \lesssim 0.5$ MeV and 
$-0.15 \lesssim \kappa_f \lesssim -0.25$ MeV 
for $^{98}$Nb, and 
$\vt \gtrsim 0.2$ MeV 
and $\kappa_f \gtrsim -0.15$ MeV for $^{100}$Nb. 
As for $^{102}$Nb, there are basically two regions 
in which the crossing of the $\ft$ curve across 
the shaded area is visible: one with 
$0 \lesssim \vt \lesssim 0.15$ MeV and 
$-0.5 \lesssim \kappa_f \lesssim -0.4$ MeV, 
and the other with 
$\vt \gtrsim 0.25$ MeV and 
$\kappa_f \gtrsim -0.25$ MeV. 
The aforementioned $\kappa_f$ values, i.e., 
$\kappa_f=-0.180$, $-0.150$, and $-0.260$ MeV, 
and $\vt=0.28$, 0.50, and 0.50 MeV 
for $^{98,100,102}$Nb, respectively, leading 
to a good agreement with the $\ft(0^+_1 \to 1^+_1)$ 
data, more or less fall into these ranges.

Figure~\ref{fig:level-nb} shows the calculated low-energy 
levels for the odd-odd $^{98,100,102}$Nb nuclei employing 
those parameters derived from the RHB-to-IBM mapping 
in Ref.~\cite{nomura2024beta}, and the optimal $\kappa_f$ 
parameter that gives reasonable agreement with the 
$\ft(0^+_1 \to 1^+_1)$ data for the $^A$Zr $\to$ $^A$Nb $\btm$ 
decay: 
$\kappa_f=-0.180$, $-0.150$, and $-0.260$ MeV, 
for $A=98$, 100, and 102, respectively. 
All the other model parameters, including $\vt$, 
are the same as those used for the mapped IBM-2 
calculations (see Table~\ref{tab:para-nb}). 
The corresponding experimental energy spectrum \cite{data}
is also included. 
An overall effect of using the optimal $\kappa_f$ strength 
parameter for $^{98,100}$Nb is to compress the energy spectrum. 
For $^{98}$Nb, the energy levels for the $4^+_1$ and $5^+_1$ 
states are lowered in the present IBFFM-2 calculation, 
and agree with the experimental data better 
than the previous calculation in 
Ref.~\cite{nomura2024beta}. 
The $3^+_1$ energy level is calculated to be very near 
the $1^+_1$ ground state in both versions of the IBFFM-2. 
It is quite at variance with experiment, 
which rather suggests this 
state to be found at the excitation energy of 737 keV. 
We note that for $^{98}$Nb spins for all the observed levels 
included in Fig.~\ref{fig:level-nb}, but for the $1^+$ 
one, have not been firmly established experimentally. 
The energy spectrum for the $^{100}$Nb obtained in the 
present work is even more compressed with respect to the 
one in the previous study \cite{nomura2024beta}. 
An improvement over the previous mapped-IBM-2 calculation 
is only seen in the energy level of the $2^+_1$ state, 
which agrees with data. 
The spins and parities of those states other than the 
$1^+$ ground state have not been 
established for $^{100}$Nb, either. 
For $^{102}$Nb, the energy spectra resulting from 
the mapped IBM-2 and the present calculation look 
strikingly similar to each other. 
This reflects the fact that the mapped IBM-2 
calculation in Ref.~\cite{nomura2024beta} already 
reproduced satisfactorily the $\ft(0^+_1 \to 1^+_1)$ 
value for the $^{102}$Zr$\to^{102}$Nb decay, and 
the $\kappa_f$ parameter employed there ($\kappa_f=-0.270$ MeV) 
is close to that optimal value 
extracted in the present study. 
Determination of the spin and parity of $^{102}$Nb 
has been under debate, 
and is only tentatively assigned to be 
$4^+$ in the NNDC database. 
But since the lowest energy state 
for which the spin and parity are 
identified is the $1^+$ state, 
we here regard the $1^+$ state 
as the ground state of $^{102}$Nb.

In Fig.~\ref{fig:level-mo}, 
we make similar comparisons to those in Fig.~\ref{fig:level-nb} 
for the IBM-2 energy spectra for the even-even 
$^{98,100,102}$Mo nuclei, since they are used as the 
even-even boson cores for the odd-odd $^{98,100,102}$Nb 
nuclei, respectively. 
Just as in the cases of the odd-odd Nb nuclei, 
a notable effect of using the optimal $\kappa_f$ 
parameter in the present analysis is 
to compress the whole energy spectrum 
with respect to the previous mapped IBM-2 
calculations for $^{98}$Mo and $^{100}$Mo. 
As for $^{102}$Mo, there is essentially no difference 
between the two IBM-2 calculations, since the 
employed $\kappa_f$ values are similar. 
In general, the new calculations reproduce the 
experimental energy spectra better than the 
previous ones for $^{98,100}$Mo.

We note that, within the mapped IBM-2 framework, 
the derived quadrupole-quadrupole strength parameter 
between unlike bosons is shown to be generally so large 
in magnitude that the resulting energy spectra, 
especially for those of the non-yrast states, 
are predicted to be considerably higher 
than the experimental ones. 
We might then attribute the too large 
derived quadrupole-quadrupole boson interaction strength 
to the properties of the EDF and/or the pairing 
properties employed in the SCMF calculations, 
upon which the mapping procedure is based. 
One immediate solution would be to adjust 
some of the parameters 
in the EDF-SCMF model so as to provide a reliable 
microscopic input to the IBM-2 that leads to an improved 
description of the low-lying states and $\beta$-decay 
properties simultaneously. 
By increasing the strength of the pairing correlations, 
for instance, the SCMF calculations yield 
the PESs that are softer in deformation variables, 
and the derived quadrupole-quadrupole strength 
in the IBM-2 Hamiltonian is expected to be smaller 
than otherwise, thus leading to a better 
description of the measured low-lying states 
as well as the $\beta$-decay properties.

It should be also noted that, for all these 
three Mo isotopes near $N=60$, 
the roles of shape coexistence 
and intruder excitations are expected to be 
significant 
\cite{heyde2011,thomas2013,thomas2016,nomura2016zr,garrett2022}, 
which is indeed reflected in the low-lying 
$0^+_2$ energy level 
found in the vicinity of the $2^+_1$ one. 
The low-lying $0^+_2$ levels 
could not be reproduced by the 
version of the IBM-2, that is adopted here and 
in Ref.~\cite{nomura2024beta}. 
The low-lying $0^+$ states and shape coexistence 
could be handled within the IBM, e.g., 
by incorporating effects of configuration mixing 
between normal and intruder states \cite{duval1981}, 
as was considered in previous mapped IBM 
calculations \cite{nomura2012sc,nomura2016zr}.

\section{Summary and conclusions\label{sec:summary}}

We have analyzed the parameter 
dependence of the calculated $\beta$-decay properties, 
as well as the low-lying states, of the neutron-rich Zr 
isotopes within the IBM-2 and IBFFM-2. 
The present analysis is a continuation of the 
preceding IBM-2 study on the $\beta$-decay properties 
in this mass region, which was based on the 
microscopic EDF framework. 
The present study aims to identify which of 
the various model parameters affect most the predictions 
on $\beta$ decay and also play a key role in 
improving the accuracy in reproducing 
the experimental data. 
The calculated $\ft$ values 
for the $\btm$ decays of the neutron-rich even-even 
$^{96-102}$Zr into odd-odd $^{96-102}$Nb nuclei 
here have been shown to exhibit consistently strong 
dependencies on the model parameters that are associated 
with the quadrupole-quadrupole interaction strength 
($\kappa_f$), and with the 
residual neutron-proton interaction of tensor 
type ($\vt$), which are involved 
in the IBFFM-2 Hamiltonian for describing the 
odd-odd daughter Nb nuclei. 
Along with the $\ft$ values, 
we have also investigated the parameter dependencies of 
the calculated excitation energies for the low-spin 
and low-lying states for the odd-odd Nb nuclei, 
and the GT strength distributions.

The problem encountered in the previous 
mapped IBM-2 calculations \cite{nomura2024beta} was that 
the measured $\btm$-decay $\ft(0^+_1 \to 1^+_1)$ values
for the $^{98}$Zr and $^{100}$Zr nuclei in particular 
were underestimated by a factor of $\approx 1.5$. 
The analyses made in the present study 
have indicated that a major cause of this 
discrepancy is the fact that the $\kappa_f$ parameter 
obtained by the mapping procedure might have been 
too large in magnitude. 
It has been indeed shown here that 
reductions in magnitude of $\kappa_f$ by approximately   
a factor of 2 would be required 
to improve the mapped-IBM-2 description 
of $\ft(0^+_1 \to 1^+_1)$ values. 
The reduced $|\kappa_f|$ value leading 
to a good agreement with the measured 
$\ft(0^+_1 \to 1^+_1)$ has been, in turn, 
employed for calculating the low-lying states 
of each nucleus, 
and it has been shown that 
the reduction of $|\kappa_f|$ improves significantly 
the description of the low-lying energy 
levels for the even-even $^{98,100}$Mo nuclei, 
which are considered as the boson cores for 
the odd-odd $^{98,100}$Nb nuclei, respectively.

The fact that the too large quadrupole-quadrupole 
interaction strength for the even-even nuclei 
was obtained in the 
mapped IBM-2 could be partly attributed to the 
properties of the EDF-SCMF calculations. 
Possible impacts of the parameters 
involved in the EDF calculations, e.g., 
pairing strengths, 
on the spectroscopic properties and fundamental 
nuclear processes, such as the single-$\beta$ and 
double-$\beta$ decays, could be investigated as 
a further step for a precise IBM description. 
In addition, as the present analysis was 
focused specifically on the Zr isotopes, it 
could be extended further to other nuclear systems. 
We could also take into account additional 
parameters in the IBM-2 and IBFFM-2 Hamiltonians 
that could have influences on the $\ft$ predictions, 
including the single-particle 
energies, and occupation probabilities. 
Sensitivities of the calculated quantities would 
be then analyzed in a larger parameter space. 
The work along these lines is in progress, and 
will be reported elsewhere.

\bibliography{refs}

\begin{thebibliography}{56}%
\makeatletter
\providecommand \@ifxundefined [1]{%
 \@ifx{#1\undefined}
}%
\providecommand \@ifnum [1]{%
 \ifnum #1\expandafter \@firstoftwo
 \else \expandafter \@secondoftwo
 \fi
}%
\providecommand \@ifx [1]{%
 \ifx #1\expandafter \@firstoftwo
 \else \expandafter \@secondoftwo
 \fi
}%
\providecommand \natexlab [1]{#1}%
\providecommand \enquote  [1]{``#1''}%
\providecommand \bibnamefont  [1]{#1}%
\providecommand \bibfnamefont [1]{#1}%
\providecommand \citenamefont [1]{#1}%
\providecommand \href@noop [0]{\@secondoftwo}%
\providecommand \href [0]{\begingroup \@sanitize@url \@href}%
\providecommand \@href[1]{\@@startlink{#1}\@@href}%
\providecommand \@@href[1]{\endgroup#1\@@endlink}%
\providecommand \@sanitize@url [0]{\catcode `\\12\catcode `\$12\catcode
  `\&12\catcode `\#12\catcode `\^12\catcode `\_12\catcode `\%12\relax}%
\providecommand \@@startlink[1]{}%
\providecommand \@@endlink[0]{}%
\providecommand \url  [0]{\begingroup\@sanitize@url \@url }%
\providecommand \@url [1]{\endgroup\@href {#1}{\urlprefix }}%
\providecommand \urlprefix  [0]{URL }%
\providecommand \Eprint [0]{\href }%
\providecommand \doibase [0]{https://doi.org/}%
\providecommand \selectlanguage [0]{\@gobble}%
\providecommand \bibinfo  [0]{\@secondoftwo}%
\providecommand \bibfield  [0]{\@secondoftwo}%
\providecommand \translation [1]{[#1]}%
\providecommand \BibitemOpen [0]{}%
\providecommand \bibitemStop [0]{}%
\providecommand \bibitemNoStop [0]{.\EOS\space}%
\providecommand \EOS [0]{\spacefactor3000\relax}%
\providecommand \BibitemShut  [1]{\csname bibitem#1\endcsname}%
\let\auto@bib@innerbib\@empty
%</preamble>
\bibitem [{\citenamefont {Dillmann}\ \emph {et~al.}(2003)\citenamefont
  {Dillmann}, \citenamefont {Kratz}, \citenamefont {W\"ohr}, \citenamefont
  {Arndt}, \citenamefont {Brown}, \citenamefont {Hoff}, \citenamefont
  {Hjorth-Jensen}, \citenamefont {K\"oster}, \citenamefont {Ostrowski},
  \citenamefont {Pfeiffer}, \citenamefont {Seweryniak}, \citenamefont
  {Shergur},\ and\ \citenamefont {Walters}}]{dillmann2003}%
  \BibitemOpen
  \bibfield  {author} {\bibinfo {author} {\bibfnamefont {I.}~\bibnamefont
  {Dillmann}}, \bibinfo {author} {\bibfnamefont {K.-L.}\ \bibnamefont {Kratz}},
  \bibinfo {author} {\bibfnamefont {A.}~\bibnamefont {W\"ohr}}, \bibinfo
  {author} {\bibfnamefont {O.}~\bibnamefont {Arndt}}, \bibinfo {author}
  {\bibfnamefont {B.~A.}\ \bibnamefont {Brown}}, \bibinfo {author}
  {\bibfnamefont {P.}~\bibnamefont {Hoff}}, \bibinfo {author} {\bibfnamefont
  {M.}~\bibnamefont {Hjorth-Jensen}}, \bibinfo {author} {\bibfnamefont
  {U.}~\bibnamefont {K\"oster}}, \bibinfo {author} {\bibfnamefont {A.~N.}\
  \bibnamefont {Ostrowski}}, \bibinfo {author} {\bibfnamefont {B.}~\bibnamefont
  {Pfeiffer}}, \bibinfo {author} {\bibfnamefont {D.}~\bibnamefont
  {Seweryniak}}, \bibinfo {author} {\bibfnamefont {J.}~\bibnamefont
  {Shergur}},\ and\ \bibinfo {author} {\bibfnamefont {W.~B.}\ \bibnamefont
  {Walters}} (\bibinfo {collaboration} {the ISOLDE Collaboration}),\ }\href
  {https://doi.org/10.1103/PhysRevLett.91.162503} {\bibfield  {journal}
  {\bibinfo  {journal} {Phys. Rev. Lett.}\ }\textbf {\bibinfo {volume} {91}},\
  \bibinfo {pages} {162503} (\bibinfo {year} {2003})}\BibitemShut {NoStop}%
\bibitem [{\citenamefont {Nishimura}\ \emph {et~al.}(2011)\citenamefont
  {Nishimura}, \citenamefont {Li}, \citenamefont {Watanabe}, \citenamefont
  {Yoshinaga}, \citenamefont {Sumikama}, \citenamefont {Tachibana},
  \citenamefont {Yamaguchi}, \citenamefont {Kurata-Nishimura}, \citenamefont
  {Lorusso}, \citenamefont {Miyashita}, \citenamefont {Odahara}, \citenamefont
  {Baba}, \citenamefont {Berryman}, \citenamefont {Blasi}, \citenamefont
  {Bracco}, \citenamefont {Camera}, \citenamefont {Chiba}, \citenamefont
  {Doornenbal}, \citenamefont {Go}, \citenamefont {Hashimoto}, \citenamefont
  {Hayakawa}, \citenamefont {Hinke}, \citenamefont {Ideguchi}, \citenamefont
  {Isobe}, \citenamefont {Ito}, \citenamefont {Jenkins}, \citenamefont
  {Kawada}, \citenamefont {Kobayashi}, \citenamefont {Kondo}, \citenamefont
  {Kr\"ucken}, \citenamefont {Kubono}, \citenamefont {Nakano}, \citenamefont
  {Ong}, \citenamefont {Ota}, \citenamefont {Podoly\'ak}, \citenamefont
  {Sakurai}, \citenamefont {Scheit}, \citenamefont {Steiger}, \citenamefont
  {Steppenbeck}, \citenamefont {Sugimoto}, \citenamefont {Takano},
  \citenamefont {Takashima}, \citenamefont {Tajiri}, \citenamefont {Teranishi},
  \citenamefont {Wakabayashi}, \citenamefont {Walker}, \citenamefont
  {Wieland},\ and\ \citenamefont {Yamaguchi}}]{nishimura2011}%
  \BibitemOpen
  \bibfield  {author} {\bibinfo {author} {\bibfnamefont {S.}~\bibnamefont
  {Nishimura}}, \bibinfo {author} {\bibfnamefont {Z.}~\bibnamefont {Li}},
  \bibinfo {author} {\bibfnamefont {H.}~\bibnamefont {Watanabe}}, \bibinfo
  {author} {\bibfnamefont {K.}~\bibnamefont {Yoshinaga}}, \bibinfo {author}
  {\bibfnamefont {T.}~\bibnamefont {Sumikama}}, \bibinfo {author}
  {\bibfnamefont {T.}~\bibnamefont {Tachibana}}, \bibinfo {author}
  {\bibfnamefont {K.}~\bibnamefont {Yamaguchi}}, \bibinfo {author}
  {\bibfnamefont {M.}~\bibnamefont {Kurata-Nishimura}}, \bibinfo {author}
  {\bibfnamefont {G.}~\bibnamefont {Lorusso}}, \bibinfo {author} {\bibfnamefont
  {Y.}~\bibnamefont {Miyashita}}, \bibinfo {author} {\bibfnamefont
  {A.}~\bibnamefont {Odahara}}, \bibinfo {author} {\bibfnamefont
  {H.}~\bibnamefont {Baba}}, \bibinfo {author} {\bibfnamefont {J.~S.}\
  \bibnamefont {Berryman}}, \bibinfo {author} {\bibfnamefont {N.}~\bibnamefont
  {Blasi}}, \bibinfo {author} {\bibfnamefont {A.}~\bibnamefont {Bracco}},
  \bibinfo {author} {\bibfnamefont {F.}~\bibnamefont {Camera}}, \bibinfo
  {author} {\bibfnamefont {J.}~\bibnamefont {Chiba}}, \bibinfo {author}
  {\bibfnamefont {P.}~\bibnamefont {Doornenbal}}, \bibinfo {author}
  {\bibfnamefont {S.}~\bibnamefont {Go}}, \bibinfo {author} {\bibfnamefont
  {T.}~\bibnamefont {Hashimoto}}, \bibinfo {author} {\bibfnamefont
  {S.}~\bibnamefont {Hayakawa}}, \bibinfo {author} {\bibfnamefont
  {C.}~\bibnamefont {Hinke}}, \bibinfo {author} {\bibfnamefont
  {E.}~\bibnamefont {Ideguchi}}, \bibinfo {author} {\bibfnamefont
  {T.}~\bibnamefont {Isobe}}, \bibinfo {author} {\bibfnamefont
  {Y.}~\bibnamefont {Ito}}, \bibinfo {author} {\bibfnamefont {D.~G.}\
  \bibnamefont {Jenkins}}, \bibinfo {author} {\bibfnamefont {Y.}~\bibnamefont
  {Kawada}}, \bibinfo {author} {\bibfnamefont {N.}~\bibnamefont {Kobayashi}},
  \bibinfo {author} {\bibfnamefont {Y.}~\bibnamefont {Kondo}}, \bibinfo
  {author} {\bibfnamefont {R.}~\bibnamefont {Kr\"ucken}}, \bibinfo {author}
  {\bibfnamefont {S.}~\bibnamefont {Kubono}}, \bibinfo {author} {\bibfnamefont
  {T.}~\bibnamefont {Nakano}}, \bibinfo {author} {\bibfnamefont {H.~J.}\
  \bibnamefont {Ong}}, \bibinfo {author} {\bibfnamefont {S.}~\bibnamefont
  {Ota}}, \bibinfo {author} {\bibfnamefont {Z.}~\bibnamefont {Podoly\'ak}},
  \bibinfo {author} {\bibfnamefont {H.}~\bibnamefont {Sakurai}}, \bibinfo
  {author} {\bibfnamefont {H.}~\bibnamefont {Scheit}}, \bibinfo {author}
  {\bibfnamefont {K.}~\bibnamefont {Steiger}}, \bibinfo {author} {\bibfnamefont
  {D.}~\bibnamefont {Steppenbeck}}, \bibinfo {author} {\bibfnamefont
  {K.}~\bibnamefont {Sugimoto}}, \bibinfo {author} {\bibfnamefont
  {S.}~\bibnamefont {Takano}}, \bibinfo {author} {\bibfnamefont
  {A.}~\bibnamefont {Takashima}}, \bibinfo {author} {\bibfnamefont
  {K.}~\bibnamefont {Tajiri}}, \bibinfo {author} {\bibfnamefont
  {T.}~\bibnamefont {Teranishi}}, \bibinfo {author} {\bibfnamefont
  {Y.}~\bibnamefont {Wakabayashi}}, \bibinfo {author} {\bibfnamefont {P.~M.}\
  \bibnamefont {Walker}}, \bibinfo {author} {\bibfnamefont {O.}~\bibnamefont
  {Wieland}},\ and\ \bibinfo {author} {\bibfnamefont {H.}~\bibnamefont
  {Yamaguchi}},\ }\href {https://doi.org/10.1103/PhysRevLett.106.052502}
  {\bibfield  {journal} {\bibinfo  {journal} {Phys. Rev. Lett.}\ }\textbf
  {\bibinfo {volume} {106}},\ \bibinfo {pages} {052502} (\bibinfo {year}
  {2011})}\BibitemShut {NoStop}%
\bibitem [{\citenamefont {Quinn}\ \emph {et~al.}(2012)\citenamefont {Quinn},
  \citenamefont {Aprahamian}, \citenamefont {Pereira}, \citenamefont {Surman},
  \citenamefont {Arndt}, \citenamefont {Baumann}, \citenamefont {Becerril},
  \citenamefont {Elliot}, \citenamefont {Estrade}, \citenamefont {Galaviz},
  \citenamefont {Ginter}, \citenamefont {Hausmann}, \citenamefont {Hennrich},
  \citenamefont {Kessler}, \citenamefont {Kratz}, \citenamefont {Lorusso},
  \citenamefont {Mantica}, \citenamefont {Matos}, \citenamefont {Montes},
  \citenamefont {Pfeiffer}, \citenamefont {Portillo}, \citenamefont {Schatz},
  \citenamefont {Schertz}, \citenamefont {Schnorrenberger}, \citenamefont
  {Smith}, \citenamefont {Stolz}, \citenamefont {Walters},\ and\ \citenamefont
  {W\"ohr}}]{quinn2012}%
  \BibitemOpen
  \bibfield  {author} {\bibinfo {author} {\bibfnamefont {M.}~\bibnamefont
  {Quinn}}, \bibinfo {author} {\bibfnamefont {A.}~\bibnamefont {Aprahamian}},
  \bibinfo {author} {\bibfnamefont {J.}~\bibnamefont {Pereira}}, \bibinfo
  {author} {\bibfnamefont {R.}~\bibnamefont {Surman}}, \bibinfo {author}
  {\bibfnamefont {O.}~\bibnamefont {Arndt}}, \bibinfo {author} {\bibfnamefont
  {T.}~\bibnamefont {Baumann}}, \bibinfo {author} {\bibfnamefont
  {A.}~\bibnamefont {Becerril}}, \bibinfo {author} {\bibfnamefont
  {T.}~\bibnamefont {Elliot}}, \bibinfo {author} {\bibfnamefont
  {A.}~\bibnamefont {Estrade}}, \bibinfo {author} {\bibfnamefont
  {D.}~\bibnamefont {Galaviz}}, \bibinfo {author} {\bibfnamefont
  {T.}~\bibnamefont {Ginter}}, \bibinfo {author} {\bibfnamefont
  {M.}~\bibnamefont {Hausmann}}, \bibinfo {author} {\bibfnamefont
  {S.}~\bibnamefont {Hennrich}}, \bibinfo {author} {\bibfnamefont
  {R.}~\bibnamefont {Kessler}}, \bibinfo {author} {\bibfnamefont {K.-L.}\
  \bibnamefont {Kratz}}, \bibinfo {author} {\bibfnamefont {G.}~\bibnamefont
  {Lorusso}}, \bibinfo {author} {\bibfnamefont {P.~F.}\ \bibnamefont
  {Mantica}}, \bibinfo {author} {\bibfnamefont {M.}~\bibnamefont {Matos}},
  \bibinfo {author} {\bibfnamefont {F.}~\bibnamefont {Montes}}, \bibinfo
  {author} {\bibfnamefont {B.}~\bibnamefont {Pfeiffer}}, \bibinfo {author}
  {\bibfnamefont {M.}~\bibnamefont {Portillo}}, \bibinfo {author}
  {\bibfnamefont {H.}~\bibnamefont {Schatz}}, \bibinfo {author} {\bibfnamefont
  {F.}~\bibnamefont {Schertz}}, \bibinfo {author} {\bibfnamefont
  {L.}~\bibnamefont {Schnorrenberger}}, \bibinfo {author} {\bibfnamefont
  {E.}~\bibnamefont {Smith}}, \bibinfo {author} {\bibfnamefont
  {A.}~\bibnamefont {Stolz}}, \bibinfo {author} {\bibfnamefont {W.~B.}\
  \bibnamefont {Walters}},\ and\ \bibinfo {author} {\bibfnamefont
  {A.}~\bibnamefont {W\"ohr}},\ }\href
  {https://doi.org/10.1103/PhysRevC.85.035807} {\bibfield  {journal} {\bibinfo
  {journal} {Phys. Rev. C}\ }\textbf {\bibinfo {volume} {85}},\ \bibinfo
  {pages} {035807} (\bibinfo {year} {2012})}\BibitemShut {NoStop}%
\bibitem [{\citenamefont {Lorusso}\ \emph {et~al.}(2015)\citenamefont
  {Lorusso}, \citenamefont {Nishimura}, \citenamefont {Xu}, \citenamefont
  {Jungclaus}, \citenamefont {Shimizu}, \citenamefont {Simpson}, \citenamefont
  {S\"oderstr\"om}, \citenamefont {Watanabe}, \citenamefont {Browne},
  \citenamefont {Doornenbal}, \citenamefont {Gey}, \citenamefont {Jung},
  \citenamefont {Meyer}, \citenamefont {Sumikama}, \citenamefont {Taprogge},
  \citenamefont {Vajta}, \citenamefont {Wu}, \citenamefont {Baba},
  \citenamefont {Benzoni}, \citenamefont {Chae}, \citenamefont {Crespi},
  \citenamefont {Fukuda}, \citenamefont {Gernh\"auser}, \citenamefont {Inabe},
  \citenamefont {Isobe}, \citenamefont {Kajino}, \citenamefont {Kameda},
  \citenamefont {Kim}, \citenamefont {Kim}, \citenamefont {Kojouharov},
  \citenamefont {Kondev}, \citenamefont {Kubo}, \citenamefont {Kurz},
  \citenamefont {Kwon}, \citenamefont {Lane}, \citenamefont {Li}, \citenamefont
  {Montaner-Piz\'a}, \citenamefont {Moschner}, \citenamefont {Naqvi},
  \citenamefont {Niikura}, \citenamefont {Nishibata}, \citenamefont {Odahara},
  \citenamefont {Orlandi}, \citenamefont {Patel}, \citenamefont {Podoly\'ak},
  \citenamefont {Sakurai}, \citenamefont {Schaffner}, \citenamefont {Schury},
  \citenamefont {Shibagaki}, \citenamefont {Steiger}, \citenamefont {Suzuki},
  \citenamefont {Takeda}, \citenamefont {Wendt}, \citenamefont {Yagi},\ and\
  \citenamefont {Yoshinaga}}]{lorusso2015}%
  \BibitemOpen
  \bibfield  {author} {\bibinfo {author} {\bibfnamefont {G.}~\bibnamefont
  {Lorusso}}, \bibinfo {author} {\bibfnamefont {S.}~\bibnamefont {Nishimura}},
  \bibinfo {author} {\bibfnamefont {Z.~Y.}\ \bibnamefont {Xu}}, \bibinfo
  {author} {\bibfnamefont {A.}~\bibnamefont {Jungclaus}}, \bibinfo {author}
  {\bibfnamefont {Y.}~\bibnamefont {Shimizu}}, \bibinfo {author} {\bibfnamefont
  {G.~S.}\ \bibnamefont {Simpson}}, \bibinfo {author} {\bibfnamefont {P.-A.}\
  \bibnamefont {S\"oderstr\"om}}, \bibinfo {author} {\bibfnamefont
  {H.}~\bibnamefont {Watanabe}}, \bibinfo {author} {\bibfnamefont
  {F.}~\bibnamefont {Browne}}, \bibinfo {author} {\bibfnamefont
  {P.}~\bibnamefont {Doornenbal}}, \bibinfo {author} {\bibfnamefont
  {G.}~\bibnamefont {Gey}}, \bibinfo {author} {\bibfnamefont {H.~S.}\
  \bibnamefont {Jung}}, \bibinfo {author} {\bibfnamefont {B.}~\bibnamefont
  {Meyer}}, \bibinfo {author} {\bibfnamefont {T.}~\bibnamefont {Sumikama}},
  \bibinfo {author} {\bibfnamefont {J.}~\bibnamefont {Taprogge}}, \bibinfo
  {author} {\bibfnamefont {Z.}~\bibnamefont {Vajta}}, \bibinfo {author}
  {\bibfnamefont {J.}~\bibnamefont {Wu}}, \bibinfo {author} {\bibfnamefont
  {H.}~\bibnamefont {Baba}}, \bibinfo {author} {\bibfnamefont {G.}~\bibnamefont
  {Benzoni}}, \bibinfo {author} {\bibfnamefont {K.~Y.}\ \bibnamefont {Chae}},
  \bibinfo {author} {\bibfnamefont {F.~C.~L.}\ \bibnamefont {Crespi}}, \bibinfo
  {author} {\bibfnamefont {N.}~\bibnamefont {Fukuda}}, \bibinfo {author}
  {\bibfnamefont {R.}~\bibnamefont {Gernh\"auser}}, \bibinfo {author}
  {\bibfnamefont {N.}~\bibnamefont {Inabe}}, \bibinfo {author} {\bibfnamefont
  {T.}~\bibnamefont {Isobe}}, \bibinfo {author} {\bibfnamefont
  {T.}~\bibnamefont {Kajino}}, \bibinfo {author} {\bibfnamefont
  {D.}~\bibnamefont {Kameda}}, \bibinfo {author} {\bibfnamefont {G.~D.}\
  \bibnamefont {Kim}}, \bibinfo {author} {\bibfnamefont {Y.-K.}\ \bibnamefont
  {Kim}}, \bibinfo {author} {\bibfnamefont {I.}~\bibnamefont {Kojouharov}},
  \bibinfo {author} {\bibfnamefont {F.~G.}\ \bibnamefont {Kondev}}, \bibinfo
  {author} {\bibfnamefont {T.}~\bibnamefont {Kubo}}, \bibinfo {author}
  {\bibfnamefont {N.}~\bibnamefont {Kurz}}, \bibinfo {author} {\bibfnamefont
  {Y.~K.}\ \bibnamefont {Kwon}}, \bibinfo {author} {\bibfnamefont {G.~J.}\
  \bibnamefont {Lane}}, \bibinfo {author} {\bibfnamefont {Z.}~\bibnamefont
  {Li}}, \bibinfo {author} {\bibfnamefont {A.}~\bibnamefont {Montaner-Piz\'a}},
  \bibinfo {author} {\bibfnamefont {K.}~\bibnamefont {Moschner}}, \bibinfo
  {author} {\bibfnamefont {F.}~\bibnamefont {Naqvi}}, \bibinfo {author}
  {\bibfnamefont {M.}~\bibnamefont {Niikura}}, \bibinfo {author} {\bibfnamefont
  {H.}~\bibnamefont {Nishibata}}, \bibinfo {author} {\bibfnamefont
  {A.}~\bibnamefont {Odahara}}, \bibinfo {author} {\bibfnamefont
  {R.}~\bibnamefont {Orlandi}}, \bibinfo {author} {\bibfnamefont
  {Z.}~\bibnamefont {Patel}}, \bibinfo {author} {\bibfnamefont
  {Z.}~\bibnamefont {Podoly\'ak}}, \bibinfo {author} {\bibfnamefont
  {H.}~\bibnamefont {Sakurai}}, \bibinfo {author} {\bibfnamefont
  {H.}~\bibnamefont {Schaffner}}, \bibinfo {author} {\bibfnamefont
  {P.}~\bibnamefont {Schury}}, \bibinfo {author} {\bibfnamefont
  {S.}~\bibnamefont {Shibagaki}}, \bibinfo {author} {\bibfnamefont
  {K.}~\bibnamefont {Steiger}}, \bibinfo {author} {\bibfnamefont
  {H.}~\bibnamefont {Suzuki}}, \bibinfo {author} {\bibfnamefont
  {H.}~\bibnamefont {Takeda}}, \bibinfo {author} {\bibfnamefont
  {A.}~\bibnamefont {Wendt}}, \bibinfo {author} {\bibfnamefont
  {A.}~\bibnamefont {Yagi}},\ and\ \bibinfo {author} {\bibfnamefont
  {K.}~\bibnamefont {Yoshinaga}},\ }\href
  {https://doi.org/10.1103/PhysRevLett.114.192501} {\bibfield  {journal}
  {\bibinfo  {journal} {Phys. Rev. Lett.}\ }\textbf {\bibinfo {volume} {114}},\
  \bibinfo {pages} {192501} (\bibinfo {year} {2015})}\BibitemShut {NoStop}%
\bibitem [{\citenamefont {Caballero-Folch}\ \emph {et~al.}(2016)\citenamefont
  {Caballero-Folch}, \citenamefont {Domingo-Pardo}, \citenamefont {Agramunt},
  \citenamefont {Algora}, \citenamefont {Ameil}, \citenamefont {Arcones},
  \citenamefont {Ayyad}, \citenamefont {Benlliure}, \citenamefont {Borzov},
  \citenamefont {Bowry}, \citenamefont {Calvi\~no}, \citenamefont {Cano-Ott},
  \citenamefont {Cort\'es}, \citenamefont {Davinson}, \citenamefont {Dillmann},
  \citenamefont {Estrade}, \citenamefont {Evdokimov}, \citenamefont
  {Faestermann}, \citenamefont {Farinon}, \citenamefont {Galaviz},
  \citenamefont {Garc\'{\i}a}, \citenamefont {Geissel}, \citenamefont
  {Gelletly}, \citenamefont {Gernh\"auser}, \citenamefont {G\'omez-Hornillos},
  \citenamefont {Guerrero}, \citenamefont {Heil}, \citenamefont {Hinke},
  \citenamefont {Kn\"obel}, \citenamefont {Kojouharov}, \citenamefont
  {Kurcewicz}, \citenamefont {Kurz}, \citenamefont {Litvinov}, \citenamefont
  {Maier}, \citenamefont {Marganiec}, \citenamefont {Marketin}, \citenamefont
  {Marta}, \citenamefont {Mart\'{\i}nez}, \citenamefont {Mart\'{\i}nez-Pinedo},
  \citenamefont {Montes}, \citenamefont {Mukha}, \citenamefont {Napoli},
  \citenamefont {Nociforo}, \citenamefont {Paradela}, \citenamefont {Pietri},
  \citenamefont {Podoly\'ak}, \citenamefont {Prochazka}, \citenamefont {Rice},
  \citenamefont {Riego}, \citenamefont {Rubio}, \citenamefont {Schaffner},
  \citenamefont {Scheidenberger}, \citenamefont {Smith}, \citenamefont {Sokol},
  \citenamefont {Steiger}, \citenamefont {Sun}, \citenamefont {Ta\'{\i}n},
  \citenamefont {Takechi}, \citenamefont {Testov}, \citenamefont {Weick},
  \citenamefont {Wilson}, \citenamefont {Winfield}, \citenamefont {Wood},
  \citenamefont {Woods},\ and\ \citenamefont {Yeremin}}]{caballero2016}%
  \BibitemOpen
  \bibfield  {author} {\bibinfo {author} {\bibfnamefont {R.}~\bibnamefont
  {Caballero-Folch}}, \bibinfo {author} {\bibfnamefont {C.}~\bibnamefont
  {Domingo-Pardo}}, \bibinfo {author} {\bibfnamefont {J.}~\bibnamefont
  {Agramunt}}, \bibinfo {author} {\bibfnamefont {A.}~\bibnamefont {Algora}},
  \bibinfo {author} {\bibfnamefont {F.}~\bibnamefont {Ameil}}, \bibinfo
  {author} {\bibfnamefont {A.}~\bibnamefont {Arcones}}, \bibinfo {author}
  {\bibfnamefont {Y.}~\bibnamefont {Ayyad}}, \bibinfo {author} {\bibfnamefont
  {J.}~\bibnamefont {Benlliure}}, \bibinfo {author} {\bibfnamefont {I.~N.}\
  \bibnamefont {Borzov}}, \bibinfo {author} {\bibfnamefont {M.}~\bibnamefont
  {Bowry}}, \bibinfo {author} {\bibfnamefont {F.}~\bibnamefont {Calvi\~no}},
  \bibinfo {author} {\bibfnamefont {D.}~\bibnamefont {Cano-Ott}}, \bibinfo
  {author} {\bibfnamefont {G.}~\bibnamefont {Cort\'es}}, \bibinfo {author}
  {\bibfnamefont {T.}~\bibnamefont {Davinson}}, \bibinfo {author}
  {\bibfnamefont {I.}~\bibnamefont {Dillmann}}, \bibinfo {author}
  {\bibfnamefont {A.}~\bibnamefont {Estrade}}, \bibinfo {author} {\bibfnamefont
  {A.}~\bibnamefont {Evdokimov}}, \bibinfo {author} {\bibfnamefont
  {T.}~\bibnamefont {Faestermann}}, \bibinfo {author} {\bibfnamefont
  {F.}~\bibnamefont {Farinon}}, \bibinfo {author} {\bibfnamefont
  {D.}~\bibnamefont {Galaviz}}, \bibinfo {author} {\bibfnamefont {A.~R.}\
  \bibnamefont {Garc\'{\i}a}}, \bibinfo {author} {\bibfnamefont
  {H.}~\bibnamefont {Geissel}}, \bibinfo {author} {\bibfnamefont
  {W.}~\bibnamefont {Gelletly}}, \bibinfo {author} {\bibfnamefont
  {R.}~\bibnamefont {Gernh\"auser}}, \bibinfo {author} {\bibfnamefont {M.~B.}\
  \bibnamefont {G\'omez-Hornillos}}, \bibinfo {author} {\bibfnamefont
  {C.}~\bibnamefont {Guerrero}}, \bibinfo {author} {\bibfnamefont
  {M.}~\bibnamefont {Heil}}, \bibinfo {author} {\bibfnamefont {C.}~\bibnamefont
  {Hinke}}, \bibinfo {author} {\bibfnamefont {R.}~\bibnamefont {Kn\"obel}},
  \bibinfo {author} {\bibfnamefont {I.}~\bibnamefont {Kojouharov}}, \bibinfo
  {author} {\bibfnamefont {J.}~\bibnamefont {Kurcewicz}}, \bibinfo {author}
  {\bibfnamefont {N.}~\bibnamefont {Kurz}}, \bibinfo {author} {\bibfnamefont
  {Y.~A.}\ \bibnamefont {Litvinov}}, \bibinfo {author} {\bibfnamefont
  {L.}~\bibnamefont {Maier}}, \bibinfo {author} {\bibfnamefont
  {J.}~\bibnamefont {Marganiec}}, \bibinfo {author} {\bibfnamefont
  {T.}~\bibnamefont {Marketin}}, \bibinfo {author} {\bibfnamefont
  {M.}~\bibnamefont {Marta}}, \bibinfo {author} {\bibfnamefont
  {T.}~\bibnamefont {Mart\'{\i}nez}}, \bibinfo {author} {\bibfnamefont
  {G.}~\bibnamefont {Mart\'{\i}nez-Pinedo}}, \bibinfo {author} {\bibfnamefont
  {F.}~\bibnamefont {Montes}}, \bibinfo {author} {\bibfnamefont
  {I.}~\bibnamefont {Mukha}}, \bibinfo {author} {\bibfnamefont {D.~R.}\
  \bibnamefont {Napoli}}, \bibinfo {author} {\bibfnamefont {C.}~\bibnamefont
  {Nociforo}}, \bibinfo {author} {\bibfnamefont {C.}~\bibnamefont {Paradela}},
  \bibinfo {author} {\bibfnamefont {S.}~\bibnamefont {Pietri}}, \bibinfo
  {author} {\bibfnamefont {Z.}~\bibnamefont {Podoly\'ak}}, \bibinfo {author}
  {\bibfnamefont {A.}~\bibnamefont {Prochazka}}, \bibinfo {author}
  {\bibfnamefont {S.}~\bibnamefont {Rice}}, \bibinfo {author} {\bibfnamefont
  {A.}~\bibnamefont {Riego}}, \bibinfo {author} {\bibfnamefont
  {B.}~\bibnamefont {Rubio}}, \bibinfo {author} {\bibfnamefont
  {H.}~\bibnamefont {Schaffner}}, \bibinfo {author} {\bibfnamefont
  {C.}~\bibnamefont {Scheidenberger}}, \bibinfo {author} {\bibfnamefont
  {K.}~\bibnamefont {Smith}}, \bibinfo {author} {\bibfnamefont
  {E.}~\bibnamefont {Sokol}}, \bibinfo {author} {\bibfnamefont
  {K.}~\bibnamefont {Steiger}}, \bibinfo {author} {\bibfnamefont
  {B.}~\bibnamefont {Sun}}, \bibinfo {author} {\bibfnamefont {J.~L.}\
  \bibnamefont {Ta\'{\i}n}}, \bibinfo {author} {\bibfnamefont {M.}~\bibnamefont
  {Takechi}}, \bibinfo {author} {\bibfnamefont {D.}~\bibnamefont {Testov}},
  \bibinfo {author} {\bibfnamefont {H.}~\bibnamefont {Weick}}, \bibinfo
  {author} {\bibfnamefont {E.}~\bibnamefont {Wilson}}, \bibinfo {author}
  {\bibfnamefont {J.~S.}\ \bibnamefont {Winfield}}, \bibinfo {author}
  {\bibfnamefont {R.}~\bibnamefont {Wood}}, \bibinfo {author} {\bibfnamefont
  {P.}~\bibnamefont {Woods}},\ and\ \bibinfo {author} {\bibfnamefont
  {A.}~\bibnamefont {Yeremin}},\ }\href
  {https://doi.org/10.1103/PhysRevLett.117.012501} {\bibfield  {journal}
  {\bibinfo  {journal} {Phys. Rev. Lett.}\ }\textbf {\bibinfo {volume} {117}},\
  \bibinfo {pages} {012501} (\bibinfo {year} {2016})}\BibitemShut {NoStop}%
\bibitem [{\citenamefont {Avignone}\ \emph {et~al.}(2008)\citenamefont
  {Avignone}, \citenamefont {Elliott},\ and\ \citenamefont
  {Engel}}]{avignone2008}%
  \BibitemOpen
  \bibfield  {author} {\bibinfo {author} {\bibfnamefont {F.~T.}\ \bibnamefont
  {Avignone}}, \bibinfo {author} {\bibfnamefont {S.~R.}\ \bibnamefont
  {Elliott}},\ and\ \bibinfo {author} {\bibfnamefont {J.}~\bibnamefont
  {Engel}},\ }\href {https://doi.org/10.1103/RevModPhys.80.481} {\bibfield
  {journal} {\bibinfo  {journal} {Rev. Mod. Phys.}\ }\textbf {\bibinfo {volume}
  {80}},\ \bibinfo {pages} {481} (\bibinfo {year} {2008})}\BibitemShut
  {NoStop}%
\bibitem [{\citenamefont {Engel}\ and\ \citenamefont
  {Men{\'{e}}ndez}(2017)}]{engel2017}%
  \BibitemOpen
  \bibfield  {author} {\bibinfo {author} {\bibfnamefont {J.}~\bibnamefont
  {Engel}}\ and\ \bibinfo {author} {\bibfnamefont {J.}~\bibnamefont
  {Men{\'{e}}ndez}},\ }\href {https://doi.org/10.1088/1361-6633/aa5bc5}
  {\bibfield  {journal} {\bibinfo  {journal} {Rep. Prog. Phys.}\ }\textbf
  {\bibinfo {volume} {80}},\ \bibinfo {pages} {046301} (\bibinfo {year}
  {2017})}\BibitemShut {NoStop}%
\bibitem [{\citenamefont {Agostini}\ \emph {et~al.}(2023)\citenamefont
  {Agostini}, \citenamefont {Benato}, \citenamefont {Detwiler}, \citenamefont
  {Men\'endez},\ and\ \citenamefont {Vissani}}]{agostini2023}%
  \BibitemOpen
  \bibfield  {author} {\bibinfo {author} {\bibfnamefont {M.}~\bibnamefont
  {Agostini}}, \bibinfo {author} {\bibfnamefont {G.}~\bibnamefont {Benato}},
  \bibinfo {author} {\bibfnamefont {J.~A.}\ \bibnamefont {Detwiler}}, \bibinfo
  {author} {\bibfnamefont {J.}~\bibnamefont {Men\'endez}},\ and\ \bibinfo
  {author} {\bibfnamefont {F.}~\bibnamefont {Vissani}},\ }\href
  {https://doi.org/10.1103/RevModPhys.95.025002} {\bibfield  {journal}
  {\bibinfo  {journal} {Rev. Mod. Phys.}\ }\textbf {\bibinfo {volume} {95}},\
  \bibinfo {pages} {025002} (\bibinfo {year} {2023})}\BibitemShut {NoStop}%
\bibitem [{\citenamefont {Langanke}\ and\ \citenamefont
  {Mart\'{\i}nez-Pinedo}(2003)}]{langanke2003}%
  \BibitemOpen
  \bibfield  {author} {\bibinfo {author} {\bibfnamefont {K.}~\bibnamefont
  {Langanke}}\ and\ \bibinfo {author} {\bibfnamefont {G.}~\bibnamefont
  {Mart\'{\i}nez-Pinedo}},\ }\href {https://doi.org/10.1103/RevModPhys.75.819}
  {\bibfield  {journal} {\bibinfo  {journal} {Rev. Mod. Phys.}\ }\textbf
  {\bibinfo {volume} {75}},\ \bibinfo {pages} {819} (\bibinfo {year}
  {2003})}\BibitemShut {NoStop}%
\bibitem [{\citenamefont {Caurier}\ \emph {et~al.}(2005)\citenamefont
  {Caurier}, \citenamefont {Mart\'{\i}nez-Pinedo}, \citenamefont {Nowacki},
  \citenamefont {Poves},\ and\ \citenamefont {Zuker}}]{caurier2005}%
  \BibitemOpen
  \bibfield  {author} {\bibinfo {author} {\bibfnamefont {E.}~\bibnamefont
  {Caurier}}, \bibinfo {author} {\bibfnamefont {G.}~\bibnamefont
  {Mart\'{\i}nez-Pinedo}}, \bibinfo {author} {\bibfnamefont {F.}~\bibnamefont
  {Nowacki}}, \bibinfo {author} {\bibfnamefont {A.}~\bibnamefont {Poves}},\
  and\ \bibinfo {author} {\bibfnamefont {A.~P.}\ \bibnamefont {Zuker}},\ }\href
  {https://doi.org/10.1103/RevModPhys.77.427} {\bibfield  {journal} {\bibinfo
  {journal} {Rev. Mod. Phys.}\ }\textbf {\bibinfo {volume} {77}},\ \bibinfo
  {pages} {427} (\bibinfo {year} {2005})}\BibitemShut {NoStop}%
\bibitem [{\citenamefont {Yoshida}\ \emph {et~al.}(2018)\citenamefont
  {Yoshida}, \citenamefont {Utsuno}, \citenamefont {Shimizu},\ and\
  \citenamefont {Otsuka}}]{syoshida2018}%
  \BibitemOpen
  \bibfield  {author} {\bibinfo {author} {\bibfnamefont {S.}~\bibnamefont
  {Yoshida}}, \bibinfo {author} {\bibfnamefont {Y.}~\bibnamefont {Utsuno}},
  \bibinfo {author} {\bibfnamefont {N.}~\bibnamefont {Shimizu}},\ and\ \bibinfo
  {author} {\bibfnamefont {T.}~\bibnamefont {Otsuka}},\ }\href
  {https://doi.org/10.1103/PhysRevC.97.054321} {\bibfield  {journal} {\bibinfo
  {journal} {Phys. Rev. C}\ }\textbf {\bibinfo {volume} {97}},\ \bibinfo
  {pages} {054321} (\bibinfo {year} {2018})}\BibitemShut {NoStop}%
\bibitem [{\citenamefont {Suzuki}\ \emph {et~al.}(2018)\citenamefont {Suzuki},
  \citenamefont {Shibagaki}, \citenamefont {Yoshida}, \citenamefont {Kajino},\
  and\ \citenamefont {Otsuka}}]{suzuki2018}%
  \BibitemOpen
  \bibfield  {author} {\bibinfo {author} {\bibfnamefont {T.}~\bibnamefont
  {Suzuki}}, \bibinfo {author} {\bibfnamefont {S.}~\bibnamefont {Shibagaki}},
  \bibinfo {author} {\bibfnamefont {T.}~\bibnamefont {Yoshida}}, \bibinfo
  {author} {\bibfnamefont {T.}~\bibnamefont {Kajino}},\ and\ \bibinfo {author}
  {\bibfnamefont {T.}~\bibnamefont {Otsuka}},\ }\href
  {https://doi.org/10.3847/1538-4357/aabfde} {\bibfield  {journal} {\bibinfo
  {journal} {Astrophys. J.}\ }\textbf {\bibinfo {volume} {859}},\ \bibinfo
  {pages} {133} (\bibinfo {year} {2018})}\BibitemShut {NoStop}%
\bibitem [{\citenamefont {\'Alvarez-Rodr\'{\i}guez}\ \emph
  {et~al.}(2004)\citenamefont {\'Alvarez-Rodr\'{\i}guez}, \citenamefont
  {Sarriguren}, \citenamefont {de~Guerra}, \citenamefont {Pacearescu},
  \citenamefont {Faessler},\ and\ \citenamefont {\ifmmode~\check{S}\else
  \v{S}\fi{}imkovic}}]{alvarez2004}%
  \BibitemOpen
  \bibfield  {author} {\bibinfo {author} {\bibfnamefont {R.}~\bibnamefont
  {\'Alvarez-Rodr\'{\i}guez}}, \bibinfo {author} {\bibfnamefont
  {P.}~\bibnamefont {Sarriguren}}, \bibinfo {author} {\bibfnamefont {E.~M.}\
  \bibnamefont {de~Guerra}}, \bibinfo {author} {\bibfnamefont {L.}~\bibnamefont
  {Pacearescu}}, \bibinfo {author} {\bibfnamefont {A.}~\bibnamefont
  {Faessler}},\ and\ \bibinfo {author} {\bibfnamefont {F.}~\bibnamefont
  {\ifmmode~\check{S}\else \v{S}\fi{}imkovic}},\ }\href
  {https://doi.org/10.1103/PhysRevC.70.064309} {\bibfield  {journal} {\bibinfo
  {journal} {Phys. Rev. C}\ }\textbf {\bibinfo {volume} {70}},\ \bibinfo
  {pages} {064309} (\bibinfo {year} {2004})}\BibitemShut {NoStop}%
\bibitem [{\citenamefont {Sarriguren}(2015)}]{sarriguren2015}%
  \BibitemOpen
  \bibfield  {author} {\bibinfo {author} {\bibfnamefont {P.}~\bibnamefont
  {Sarriguren}},\ }\href {https://doi.org/10.1103/PhysRevC.91.044304}
  {\bibfield  {journal} {\bibinfo  {journal} {Phys. Rev. C}\ }\textbf {\bibinfo
  {volume} {91}},\ \bibinfo {pages} {044304} (\bibinfo {year}
  {2015})}\BibitemShut {NoStop}%
\bibitem [{\citenamefont {Boillos}\ and\ \citenamefont
  {Sarriguren}(2015)}]{boillos2015}%
  \BibitemOpen
  \bibfield  {author} {\bibinfo {author} {\bibfnamefont {J.~M.}\ \bibnamefont
  {Boillos}}\ and\ \bibinfo {author} {\bibfnamefont {P.}~\bibnamefont
  {Sarriguren}},\ }\href {https://doi.org/10.1103/PhysRevC.91.034311}
  {\bibfield  {journal} {\bibinfo  {journal} {Phys. Rev. C}\ }\textbf {\bibinfo
  {volume} {91}},\ \bibinfo {pages} {034311} (\bibinfo {year}
  {2015})}\BibitemShut {NoStop}%
\bibitem [{\citenamefont {Pirinen}\ and\ \citenamefont
  {Suhonen}(2015)}]{pirinen2015}%
  \BibitemOpen
  \bibfield  {author} {\bibinfo {author} {\bibfnamefont {P.}~\bibnamefont
  {Pirinen}}\ and\ \bibinfo {author} {\bibfnamefont {J.}~\bibnamefont
  {Suhonen}},\ }\href {https://doi.org/10.1103/PhysRevC.91.054309} {\bibfield
  {journal} {\bibinfo  {journal} {Phys. Rev. C}\ }\textbf {\bibinfo {volume}
  {91}},\ \bibinfo {pages} {054309} (\bibinfo {year} {2015})}\BibitemShut
  {NoStop}%
\bibitem [{\citenamefont {\ifmmode~\check{S}\else \v{S}\fi{}imkovic}\ \emph
  {et~al.}(2013)\citenamefont {\ifmmode~\check{S}\else \v{S}\fi{}imkovic},
  \citenamefont {Rodin}, \citenamefont {Faessler},\ and\ \citenamefont
  {Vogel}}]{simkovic2013}%
  \BibitemOpen
  \bibfield  {author} {\bibinfo {author} {\bibfnamefont {F.}~\bibnamefont
  {\ifmmode~\check{S}\else \v{S}\fi{}imkovic}}, \bibinfo {author}
  {\bibfnamefont {V.}~\bibnamefont {Rodin}}, \bibinfo {author} {\bibfnamefont
  {A.}~\bibnamefont {Faessler}},\ and\ \bibinfo {author} {\bibfnamefont
  {P.}~\bibnamefont {Vogel}},\ }\href
  {https://doi.org/10.1103/PhysRevC.87.045501} {\bibfield  {journal} {\bibinfo
  {journal} {Phys. Rev. C}\ }\textbf {\bibinfo {volume} {87}},\ \bibinfo
  {pages} {045501} (\bibinfo {year} {2013})}\BibitemShut {NoStop}%
\bibitem [{\citenamefont {Mustonen}\ and\ \citenamefont
  {Engel}(2016)}]{mustonen2016}%
  \BibitemOpen
  \bibfield  {author} {\bibinfo {author} {\bibfnamefont {M.~T.}\ \bibnamefont
  {Mustonen}}\ and\ \bibinfo {author} {\bibfnamefont {J.}~\bibnamefont
  {Engel}},\ }\href {https://doi.org/10.1103/PhysRevC.93.014304} {\bibfield
  {journal} {\bibinfo  {journal} {Phys. Rev. C}\ }\textbf {\bibinfo {volume}
  {93}},\ \bibinfo {pages} {014304} (\bibinfo {year} {2016})}\BibitemShut
  {NoStop}%
\bibitem [{\citenamefont {Suhonen}(2017)}]{suhonen2017}%
  \BibitemOpen
  \bibfield  {author} {\bibinfo {author} {\bibfnamefont {J.~T.}\ \bibnamefont
  {Suhonen}},\ }\href {https://doi.org/10.3389/fphy.2017.00055} {\bibfield
  {journal} {\bibinfo  {journal} {Frontiers Phys.}\ }\textbf {\bibinfo {volume}
  {5}},\ \bibinfo {pages} {55} (\bibinfo {year} {2017})}\BibitemShut {NoStop}%
\bibitem [{\citenamefont {Ravli\ifmmode~\acute{c}\else \'{c}\fi{}}\ \emph
  {et~al.}(2021)\citenamefont {Ravli\ifmmode~\acute{c}\else \'{c}\fi{}},
  \citenamefont {Y\"uksel}, \citenamefont {Niu},\ and\ \citenamefont
  {Paar}}]{ravlic2021}%
  \BibitemOpen
  \bibfield  {author} {\bibinfo {author} {\bibfnamefont {A.}~\bibnamefont
  {Ravli\ifmmode~\acute{c}\else \'{c}\fi{}}}, \bibinfo {author} {\bibfnamefont
  {E.}~\bibnamefont {Y\"uksel}}, \bibinfo {author} {\bibfnamefont {Y.~F.}\
  \bibnamefont {Niu}},\ and\ \bibinfo {author} {\bibfnamefont {N.}~\bibnamefont
  {Paar}},\ }\href {https://doi.org/10.1103/PhysRevC.104.054318} {\bibfield
  {journal} {\bibinfo  {journal} {Phys. Rev. C}\ }\textbf {\bibinfo {volume}
  {104}},\ \bibinfo {pages} {054318} (\bibinfo {year} {2021})}\BibitemShut
  {NoStop}%
\bibitem [{\citenamefont {Yoshida}\ \emph {et~al.}(2023)\citenamefont
  {Yoshida}, \citenamefont {Niu},\ and\ \citenamefont {Minato}}]{yoshida2023}%
  \BibitemOpen
  \bibfield  {author} {\bibinfo {author} {\bibfnamefont {K.}~\bibnamefont
  {Yoshida}}, \bibinfo {author} {\bibfnamefont {Y.}~\bibnamefont {Niu}},\ and\
  \bibinfo {author} {\bibfnamefont {F.}~\bibnamefont {Minato}},\ }\href
  {https://doi.org/10.1103/PhysRevC.108.034305} {\bibfield  {journal} {\bibinfo
   {journal} {Phys. Rev. C}\ }\textbf {\bibinfo {volume} {108}},\ \bibinfo
  {pages} {034305} (\bibinfo {year} {2023})}\BibitemShut {NoStop}%
\bibitem [{\citenamefont {Navr\'atil}\ and\ \citenamefont
  {Dobe}(1988)}]{navratil1988}%
  \BibitemOpen
  \bibfield  {author} {\bibinfo {author} {\bibfnamefont {P.}~\bibnamefont
  {Navr\'atil}}\ and\ \bibinfo {author} {\bibfnamefont {J.}~\bibnamefont
  {Dobe}},\ }\href {https://doi.org/10.1103/PhysRevC.37.2126} {\bibfield
  {journal} {\bibinfo  {journal} {Phys. Rev. C}\ }\textbf {\bibinfo {volume}
  {37}},\ \bibinfo {pages} {2126} (\bibinfo {year} {1988})}\BibitemShut
  {NoStop}%
\bibitem [{\citenamefont {Dellagiacoma}(1988)}]{dellagiacoma1988phdthesis}%
  \BibitemOpen
  \bibfield  {author} {\bibinfo {author} {\bibfnamefont {F.}~\bibnamefont
  {Dellagiacoma}},\ }\emph {\bibinfo {title} {Beta decay of odd mass nuclei in
  the interacting boson-fermion model}},\ \href@noop {} {Ph.D. thesis},\
  \bibinfo  {school} {Yale University} (\bibinfo {year} {1988})\BibitemShut
  {NoStop}%
\bibitem [{\citenamefont {Dellagiacoma}\ and\ \citenamefont
  {Iachello}(1989)}]{DELLAGIACOMA1989}%
  \BibitemOpen
  \bibfield  {author} {\bibinfo {author} {\bibfnamefont {F.}~\bibnamefont
  {Dellagiacoma}}\ and\ \bibinfo {author} {\bibfnamefont {F.}~\bibnamefont
  {Iachello}},\ }\href
  {https://doi.org/https://doi.org/10.1016/0370-2693(89)91434-2} {\bibfield
  {journal} {\bibinfo  {journal} {Phys. Lett. B}\ }\textbf {\bibinfo {volume}
  {218}},\ \bibinfo {pages} {399 } (\bibinfo {year} {1989})}\BibitemShut
  {NoStop}%
\bibitem [{\citenamefont {Brant}\ \emph {et~al.}(2004)\citenamefont {Brant},
  \citenamefont {Yoshida},\ and\ \citenamefont {Zuffi}}]{brant2004}%
  \BibitemOpen
  \bibfield  {author} {\bibinfo {author} {\bibfnamefont {S.}~\bibnamefont
  {Brant}}, \bibinfo {author} {\bibfnamefont {N.}~\bibnamefont {Yoshida}},\
  and\ \bibinfo {author} {\bibfnamefont {L.}~\bibnamefont {Zuffi}},\ }\href
  {https://doi.org/10.1103/PhysRevC.70.054301} {\bibfield  {journal} {\bibinfo
  {journal} {Phys. Rev. C}\ }\textbf {\bibinfo {volume} {70}},\ \bibinfo
  {pages} {054301} (\bibinfo {year} {2004})}\BibitemShut {NoStop}%
\bibitem [{\citenamefont {Brant}\ \emph {et~al.}(2006)\citenamefont {Brant},
  \citenamefont {Yoshida},\ and\ \citenamefont {Zuffi}}]{brant2006}%
  \BibitemOpen
  \bibfield  {author} {\bibinfo {author} {\bibfnamefont {S.}~\bibnamefont
  {Brant}}, \bibinfo {author} {\bibfnamefont {N.}~\bibnamefont {Yoshida}},\
  and\ \bibinfo {author} {\bibfnamefont {L.}~\bibnamefont {Zuffi}},\ }\href
  {https://doi.org/10.1103/PhysRevC.74.024303} {\bibfield  {journal} {\bibinfo
  {journal} {Phys. Rev. C}\ }\textbf {\bibinfo {volume} {74}},\ \bibinfo
  {pages} {024303} (\bibinfo {year} {2006})}\BibitemShut {NoStop}%
\bibitem [{\citenamefont {Yoshida}\ and\ \citenamefont
  {Iachello}(2013)}]{yoshida2013}%
  \BibitemOpen
  \bibfield  {author} {\bibinfo {author} {\bibfnamefont {N.}~\bibnamefont
  {Yoshida}}\ and\ \bibinfo {author} {\bibfnamefont {F.}~\bibnamefont
  {Iachello}},\ }\href {https://doi.org/10.1093/ptep/ptt007} {\bibfield
  {journal} {\bibinfo  {journal} {Prog. Theor. Exp. Phys.}\ }\textbf {\bibinfo
  {volume} {2013}},\ \bibinfo {pages} {043D01} (\bibinfo {year}
  {2013})}\BibitemShut {NoStop}%
\bibitem [{\citenamefont {Mardones}\ \emph {et~al.}(2016)\citenamefont
  {Mardones}, \citenamefont {Barea}, \citenamefont {Alonso},\ and\
  \citenamefont {Arias}}]{mardones2016}%
  \BibitemOpen
  \bibfield  {author} {\bibinfo {author} {\bibfnamefont {E.}~\bibnamefont
  {Mardones}}, \bibinfo {author} {\bibfnamefont {J.}~\bibnamefont {Barea}},
  \bibinfo {author} {\bibfnamefont {C.~E.}\ \bibnamefont {Alonso}},\ and\
  \bibinfo {author} {\bibfnamefont {J.~M.}\ \bibnamefont {Arias}},\ }\href
  {https://doi.org/10.1103/PhysRevC.93.034332} {\bibfield  {journal} {\bibinfo
  {journal} {Phys. Rev. C}\ }\textbf {\bibinfo {volume} {93}},\ \bibinfo
  {pages} {034332} (\bibinfo {year} {2016})}\BibitemShut {NoStop}%
\bibitem [{\citenamefont {Nomura}\ \emph
  {et~al.}(2020{\natexlab{a}})\citenamefont {Nomura}, \citenamefont
  {Rodr\'{\i}guez-Guzm\'an},\ and\ \citenamefont {Robledo}}]{nomura2020beta-1}%
  \BibitemOpen
  \bibfield  {author} {\bibinfo {author} {\bibfnamefont {K.}~\bibnamefont
  {Nomura}}, \bibinfo {author} {\bibfnamefont {R.}~\bibnamefont
  {Rodr\'{\i}guez-Guzm\'an}},\ and\ \bibinfo {author} {\bibfnamefont {L.~M.}\
  \bibnamefont {Robledo}},\ }\href
  {https://doi.org/10.1103/PhysRevC.101.024311} {\bibfield  {journal} {\bibinfo
   {journal} {Phys. Rev. C}\ }\textbf {\bibinfo {volume} {101}},\ \bibinfo
  {pages} {024311} (\bibinfo {year} {2020}{\natexlab{a}})}\BibitemShut
  {NoStop}%
\bibitem [{\citenamefont {Nomura}\ \emph
  {et~al.}(2020{\natexlab{b}})\citenamefont {Nomura}, \citenamefont
  {Rodr\'{\i}guez-Guzm\'an},\ and\ \citenamefont {Robledo}}]{nomura2020beta-2}%
  \BibitemOpen
  \bibfield  {author} {\bibinfo {author} {\bibfnamefont {K.}~\bibnamefont
  {Nomura}}, \bibinfo {author} {\bibfnamefont {R.}~\bibnamefont
  {Rodr\'{\i}guez-Guzm\'an}},\ and\ \bibinfo {author} {\bibfnamefont {L.~M.}\
  \bibnamefont {Robledo}},\ }\href
  {https://doi.org/10.1103/PhysRevC.101.044318} {\bibfield  {journal} {\bibinfo
   {journal} {Phys. Rev. C}\ }\textbf {\bibinfo {volume} {101}},\ \bibinfo
  {pages} {044318} (\bibinfo {year} {2020}{\natexlab{b}})}\BibitemShut
  {NoStop}%
\bibitem [{\citenamefont {Ferretti}\ \emph {et~al.}(2020)\citenamefont
  {Ferretti}, \citenamefont {Kotila}, \citenamefont {Vsevolodovna},\ and\
  \citenamefont {Santopinto}}]{ferretti2020}%
  \BibitemOpen
  \bibfield  {author} {\bibinfo {author} {\bibfnamefont {J.}~\bibnamefont
  {Ferretti}}, \bibinfo {author} {\bibfnamefont {J.}~\bibnamefont {Kotila}},
  \bibinfo {author} {\bibfnamefont {R.~I. M.~n.}\ \bibnamefont
  {Vsevolodovna}},\ and\ \bibinfo {author} {\bibfnamefont {E.}~\bibnamefont
  {Santopinto}},\ }\href {https://doi.org/10.1103/PhysRevC.102.054329}
  {\bibfield  {journal} {\bibinfo  {journal} {Phys. Rev. C}\ }\textbf {\bibinfo
  {volume} {102}},\ \bibinfo {pages} {054329} (\bibinfo {year}
  {2020})}\BibitemShut {NoStop}%
\bibitem [{\citenamefont {Nomura}(2022)}]{nomura2022beta-ge}%
  \BibitemOpen
  \bibfield  {author} {\bibinfo {author} {\bibfnamefont {K.}~\bibnamefont
  {Nomura}},\ }\href {https://doi.org/10.1103/PhysRevC.105.044306} {\bibfield
  {journal} {\bibinfo  {journal} {Phys. Rev. C}\ }\textbf {\bibinfo {volume}
  {105}},\ \bibinfo {pages} {044306} (\bibinfo {year} {2022})}\BibitemShut
  {NoStop}%
\bibitem [{\citenamefont {Nomura}\ \emph {et~al.}(2022)\citenamefont {Nomura},
  \citenamefont {Lotina}, \citenamefont {Rodr\'{\i}guez-Guzm\'an},\ and\
  \citenamefont {Robledo}}]{nomura2022beta-rh}%
  \BibitemOpen
  \bibfield  {author} {\bibinfo {author} {\bibfnamefont {K.}~\bibnamefont
  {Nomura}}, \bibinfo {author} {\bibfnamefont {L.}~\bibnamefont {Lotina}},
  \bibinfo {author} {\bibfnamefont {R.}~\bibnamefont
  {Rodr\'{\i}guez-Guzm\'an}},\ and\ \bibinfo {author} {\bibfnamefont {L.~M.}\
  \bibnamefont {Robledo}},\ }\href
  {https://doi.org/10.1103/PhysRevC.106.064304} {\bibfield  {journal} {\bibinfo
   {journal} {Phys. Rev. C}\ }\textbf {\bibinfo {volume} {106}},\ \bibinfo
  {pages} {064304} (\bibinfo {year} {2022})}\BibitemShut {NoStop}%
\bibitem [{\citenamefont {Maga\~na Vsevolodovna}\ \emph
  {et~al.}(2022)\citenamefont {Maga\~na Vsevolodovna}, \citenamefont
  {Santopinto},\ and\ \citenamefont {Bijker}}]{Vsevolodovna2022}%
  \BibitemOpen
  \bibfield  {author} {\bibinfo {author} {\bibfnamefont {R.~I.}\ \bibnamefont
  {Maga\~na Vsevolodovna}}, \bibinfo {author} {\bibfnamefont {E.}~\bibnamefont
  {Santopinto}},\ and\ \bibinfo {author} {\bibfnamefont {R.}~\bibnamefont
  {Bijker}},\ }\href {https://doi.org/10.1103/PhysRevC.106.044307} {\bibfield
  {journal} {\bibinfo  {journal} {Phys. Rev. C}\ }\textbf {\bibinfo {volume}
  {106}},\ \bibinfo {pages} {044307} (\bibinfo {year} {2022})}\BibitemShut
  {NoStop}%
\bibitem [{\citenamefont {Nomura}(2024)}]{nomura2024beta}%
  \BibitemOpen
  \bibfield  {author} {\bibinfo {author} {\bibfnamefont {K.}~\bibnamefont
  {Nomura}},\ }\href {https://doi.org/10.1103/PhysRevC.109.034319} {\bibfield
  {journal} {\bibinfo  {journal} {Phys. Rev. C}\ }\textbf {\bibinfo {volume}
  {109}},\ \bibinfo {pages} {034319} (\bibinfo {year} {2024})}\BibitemShut
  {NoStop}%
\bibitem [{\citenamefont {Vretenar}\ \emph {et~al.}(2005)\citenamefont
  {Vretenar}, \citenamefont {Afanasjev}, \citenamefont {Lalazissis},\ and\
  \citenamefont {Ring}}]{vretenar2005}%
  \BibitemOpen
  \bibfield  {author} {\bibinfo {author} {\bibfnamefont {D.}~\bibnamefont
  {Vretenar}}, \bibinfo {author} {\bibfnamefont {A.~V.}\ \bibnamefont
  {Afanasjev}}, \bibinfo {author} {\bibfnamefont {G.~A.}\ \bibnamefont
  {Lalazissis}},\ and\ \bibinfo {author} {\bibfnamefont {P.}~\bibnamefont
  {Ring}},\ }\href {https://doi.org/10.1016/j.physrep.2004.10.001} {\bibfield
  {journal} {\bibinfo  {journal} {Phys. Rep.}\ }\textbf {\bibinfo {volume}
  {409}},\ \bibinfo {pages} {101 } (\bibinfo {year} {2005})}\BibitemShut
  {NoStop}%
\bibitem [{\citenamefont {Nik\ifmmode \check{s}\else
  \v{s}\fi{}i\ifmmode~\acute{c}\else \'{c}\fi{}}\ \emph
  {et~al.}(2011)\citenamefont {Nik\ifmmode \check{s}\else
  \v{s}\fi{}i\ifmmode~\acute{c}\else \'{c}\fi{}}, \citenamefont {Vretenar},\
  and\ \citenamefont {Ring}}]{niksic2011}%
  \BibitemOpen
  \bibfield  {author} {\bibinfo {author} {\bibfnamefont {T.}~\bibnamefont
  {Nik\ifmmode \check{s}\else \v{s}\fi{}i\ifmmode~\acute{c}\else \'{c}\fi{}}},
  \bibinfo {author} {\bibfnamefont {D.}~\bibnamefont {Vretenar}},\ and\
  \bibinfo {author} {\bibfnamefont {P.}~\bibnamefont {Ring}},\ }\href
  {https://doi.org/10.1016/j.ppnp.2011.01.055} {\bibfield  {journal} {\bibinfo
  {journal} {Prog. Part. Nucl. Phys.}\ }\textbf {\bibinfo {volume} {66}},\
  \bibinfo {pages} {519} (\bibinfo {year} {2011})}\BibitemShut {NoStop}%
\bibitem [{\citenamefont {Nik\ifmmode \check{s}\else
  \v{s}\fi{}i\ifmmode~\acute{c}\else \'{c}\fi{}}\ \emph
  {et~al.}(2008)\citenamefont {Nik\ifmmode \check{s}\else
  \v{s}\fi{}i\ifmmode~\acute{c}\else \'{c}\fi{}}, \citenamefont {Vretenar},\
  and\ \citenamefont {Ring}}]{DDPC1}%
  \BibitemOpen
  \bibfield  {author} {\bibinfo {author} {\bibfnamefont {T.}~\bibnamefont
  {Nik\ifmmode \check{s}\else \v{s}\fi{}i\ifmmode~\acute{c}\else \'{c}\fi{}}},
  \bibinfo {author} {\bibfnamefont {D.}~\bibnamefont {Vretenar}},\ and\
  \bibinfo {author} {\bibfnamefont {P.}~\bibnamefont {Ring}},\ }\href
  {https://doi.org/10.1103/PhysRevC.78.034318} {\bibfield  {journal} {\bibinfo
  {journal} {Phys. Rev. C}\ }\textbf {\bibinfo {volume} {78}},\ \bibinfo
  {pages} {034318} (\bibinfo {year} {2008})}\BibitemShut {NoStop}%
\bibitem [{\citenamefont {Tian}\ \emph {et~al.}(2009)\citenamefont {Tian},
  \citenamefont {Ma},\ and\ \citenamefont {Ring}}]{tian2009}%
  \BibitemOpen
  \bibfield  {author} {\bibinfo {author} {\bibfnamefont {Y.}~\bibnamefont
  {Tian}}, \bibinfo {author} {\bibfnamefont {Z.~Y.}\ \bibnamefont {Ma}},\ and\
  \bibinfo {author} {\bibfnamefont {P.}~\bibnamefont {Ring}},\ }\href
  {https://doi.org/10.1016/j.physletb.2009.04.067} {\bibfield  {journal}
  {\bibinfo  {journal} {Phys. Lett. B}\ }\textbf {\bibinfo {volume} {676}},\
  \bibinfo {pages} {44 } (\bibinfo {year} {2009})}\BibitemShut {NoStop}%
\bibitem [{\citenamefont {Ginocchio}\ and\ \citenamefont
  {Kirson}(1980)}]{ginocchio1980}%
  \BibitemOpen
  \bibfield  {author} {\bibinfo {author} {\bibfnamefont {J.~N.}\ \bibnamefont
  {Ginocchio}}\ and\ \bibinfo {author} {\bibfnamefont {M.~W.}\ \bibnamefont
  {Kirson}},\ }\href {https://doi.org/10.1016/0375-9474(80)90387-5} {\bibfield
  {journal} {\bibinfo  {journal} {Nucl. Phys. A}\ }\textbf {\bibinfo {volume}
  {350}},\ \bibinfo {pages} {31} (\bibinfo {year} {1980})}\BibitemShut
  {NoStop}%
\bibitem [{\citenamefont {Brant}\ \emph {et~al.}(1984)\citenamefont {Brant},
  \citenamefont {Paar},\ and\ \citenamefont {Vretenar}}]{brant1984}%
  \BibitemOpen
  \bibfield  {author} {\bibinfo {author} {\bibfnamefont {S.}~\bibnamefont
  {Brant}}, \bibinfo {author} {\bibfnamefont {V.}~\bibnamefont {Paar}},\ and\
  \bibinfo {author} {\bibfnamefont {D.}~\bibnamefont {Vretenar}},\ }\href
  {https://doi.org/10.1007/BF01412551} {\bibfield  {journal} {\bibinfo
  {journal} {Z. Phys. A}\ }\textbf {\bibinfo {volume} {319}},\ \bibinfo {pages}
  {355} (\bibinfo {year} {1984})}\BibitemShut {NoStop}%
\bibitem [{\citenamefont {Iachello}\ and\ \citenamefont {{Van
  Isacker}}(1991)}]{IBFM}%
  \BibitemOpen
  \bibfield  {author} {\bibinfo {author} {\bibfnamefont {F.}~\bibnamefont
  {Iachello}}\ and\ \bibinfo {author} {\bibfnamefont {P.}~\bibnamefont {{Van
  Isacker}}},\ }\href@noop {} {\emph {\bibinfo {title} {The interacting
  boson-fermion model}}}\ (\bibinfo  {publisher} {Cambridge University Press,
  Cambridge},\ \bibinfo {year} {1991})\BibitemShut {NoStop}%
\bibitem [{\citenamefont {Heyde}\ and\ \citenamefont {Wood}(2011)}]{heyde2011}%
  \BibitemOpen
  \bibfield  {author} {\bibinfo {author} {\bibfnamefont {K.}~\bibnamefont
  {Heyde}}\ and\ \bibinfo {author} {\bibfnamefont {J.~L.}\ \bibnamefont
  {Wood}},\ }\href {https://doi.org/10.1103/RevModPhys.83.1467} {\bibfield
  {journal} {\bibinfo  {journal} {Rev. Mod. Phys.}\ }\textbf {\bibinfo {volume}
  {83}},\ \bibinfo {pages} {1467} (\bibinfo {year} {2011})}\BibitemShut
  {NoStop}%
\bibitem [{\citenamefont {Iachello}\ and\ \citenamefont {Arima}(1987)}]{IBM}%
  \BibitemOpen
  \bibfield  {author} {\bibinfo {author} {\bibfnamefont {F.}~\bibnamefont
  {Iachello}}\ and\ \bibinfo {author} {\bibfnamefont {A.}~\bibnamefont
  {Arima}},\ }\href@noop {} {\emph {\bibinfo {title} {The interacting boson
  model}}}\ (\bibinfo  {publisher} {Cambridge University Press, Cambridge},\
  \bibinfo {year} {1987})\BibitemShut {NoStop}%
\bibitem [{\citenamefont {Otsuka}\ \emph
  {et~al.}(1978{\natexlab{a}})\citenamefont {Otsuka}, \citenamefont {Arima},
  \citenamefont {Iachello},\ and\ \citenamefont {Talmi}}]{OAIT}%
  \BibitemOpen
  \bibfield  {author} {\bibinfo {author} {\bibfnamefont {T.}~\bibnamefont
  {Otsuka}}, \bibinfo {author} {\bibfnamefont {A.}~\bibnamefont {Arima}},
  \bibinfo {author} {\bibfnamefont {F.}~\bibnamefont {Iachello}},\ and\
  \bibinfo {author} {\bibfnamefont {I.}~\bibnamefont {Talmi}},\ }\href
  {https://doi.org/10.1016/0370-2693(78)90260-5} {\bibfield  {journal}
  {\bibinfo  {journal} {Phys. Lett. B}\ }\textbf {\bibinfo {volume} {76}},\
  \bibinfo {pages} {139 } (\bibinfo {year} {1978}{\natexlab{a}})}\BibitemShut
  {NoStop}%
\bibitem [{\citenamefont {Otsuka}\ \emph
  {et~al.}(1978{\natexlab{b}})\citenamefont {Otsuka}, \citenamefont {Arima},\
  and\ \citenamefont {Iachello}}]{OAI}%
  \BibitemOpen
  \bibfield  {author} {\bibinfo {author} {\bibfnamefont {T.}~\bibnamefont
  {Otsuka}}, \bibinfo {author} {\bibfnamefont {A.}~\bibnamefont {Arima}},\ and\
  \bibinfo {author} {\bibfnamefont {F.}~\bibnamefont {Iachello}},\ }\href
  {https://doi.org/10.1016/0375-9474(78)90532-8} {\bibfield  {journal}
  {\bibinfo  {journal} {Nucl. Phys. A}\ }\textbf {\bibinfo {volume} {309}},\
  \bibinfo {pages} {1} (\bibinfo {year} {1978}{\natexlab{b}})}\BibitemShut
  {NoStop}%
\bibitem [{\citenamefont {Scholten}(1985)}]{scholten1985}%
  \BibitemOpen
  \bibfield  {author} {\bibinfo {author} {\bibfnamefont {O.}~\bibnamefont
  {Scholten}},\ }\href
  {https://doi.org/https://doi.org/10.1016/0146-6410(85)90054-7} {\bibfield
  {journal} {\bibinfo  {journal} {Prog. Part. Nucl. Phys.}\ }\textbf {\bibinfo
  {volume} {14}},\ \bibinfo {pages} {189} (\bibinfo {year} {1985})}\BibitemShut
  {NoStop}%
\bibitem [{\citenamefont {Mizusaki}\ and\ \citenamefont
  {Otsuka}(1996)}]{mizusaki1996}%
  \BibitemOpen
  \bibfield  {author} {\bibinfo {author} {\bibfnamefont {T.}~\bibnamefont
  {Mizusaki}}\ and\ \bibinfo {author} {\bibfnamefont {T.}~\bibnamefont
  {Otsuka}},\ }\href {https://doi.org/10.1143/PTPS.125.97} {\bibfield
  {journal} {\bibinfo  {journal} {Prog. Theor. Phys. Suppl.}\ }\textbf
  {\bibinfo {volume} {125}},\ \bibinfo {pages} {97} (\bibinfo {year}
  {1996})}\BibitemShut {NoStop}%
\bibitem [{\citenamefont {Nomura}\ \emph {et~al.}(2008)\citenamefont {Nomura},
  \citenamefont {Shimizu},\ and\ \citenamefont {Otsuka}}]{nomura2008}%
  \BibitemOpen
  \bibfield  {author} {\bibinfo {author} {\bibfnamefont {K.}~\bibnamefont
  {Nomura}}, \bibinfo {author} {\bibfnamefont {N.}~\bibnamefont {Shimizu}},\
  and\ \bibinfo {author} {\bibfnamefont {T.}~\bibnamefont {Otsuka}},\ }\href
  {https://doi.org/10.1103/PhysRevLett.101.142501} {\bibfield  {journal}
  {\bibinfo  {journal} {Phys. Rev. Lett.}\ }\textbf {\bibinfo {volume} {101}},\
  \bibinfo {pages} {142501} (\bibinfo {year} {2008})}\BibitemShut {NoStop}%
\bibitem [{\citenamefont {{Brookhaven National Nuclear Data Center}}()}]{data}%
  \BibitemOpen
  \bibfield  {author} {\bibinfo {author} {\bibnamefont {{Brookhaven National
  Nuclear Data Center}}},\ }\href@noop {} {}\bibinfo {howpublished}
  {{http://www.nndc.bnl.gov}}\BibitemShut {NoStop}%
\bibitem [{\citenamefont {Thomas}\ \emph {et~al.}(2013)\citenamefont {Thomas},
  \citenamefont {Nomura}, \citenamefont {Werner}, \citenamefont {Ahn},
  \citenamefont {Cooper}, \citenamefont {Duckwitz}, \citenamefont {Hinton},
  \citenamefont {Ilie}, \citenamefont {Jolie}, \citenamefont {Petkov},\ and\
  \citenamefont {Radeck}}]{thomas2013}%
  \BibitemOpen
  \bibfield  {author} {\bibinfo {author} {\bibfnamefont {T.}~\bibnamefont
  {Thomas}}, \bibinfo {author} {\bibfnamefont {K.}~\bibnamefont {Nomura}},
  \bibinfo {author} {\bibfnamefont {V.}~\bibnamefont {Werner}}, \bibinfo
  {author} {\bibfnamefont {T.}~\bibnamefont {Ahn}}, \bibinfo {author}
  {\bibfnamefont {N.}~\bibnamefont {Cooper}}, \bibinfo {author} {\bibfnamefont
  {H.}~\bibnamefont {Duckwitz}}, \bibinfo {author} {\bibfnamefont
  {M.}~\bibnamefont {Hinton}}, \bibinfo {author} {\bibfnamefont
  {G.}~\bibnamefont {Ilie}}, \bibinfo {author} {\bibfnamefont {J.}~\bibnamefont
  {Jolie}}, \bibinfo {author} {\bibfnamefont {P.}~\bibnamefont {Petkov}},\ and\
  \bibinfo {author} {\bibfnamefont {D.}~\bibnamefont {Radeck}},\ }\href
  {https://doi.org/10.1103/PhysRevC.88.044305} {\bibfield  {journal} {\bibinfo
  {journal} {Phys. Rev. C}\ }\textbf {\bibinfo {volume} {88}},\ \bibinfo
  {pages} {044305} (\bibinfo {year} {2013})}\BibitemShut {NoStop}%
\bibitem [{\citenamefont {Thomas}\ \emph {et~al.}(2016)\citenamefont {Thomas},
  \citenamefont {Werner}, \citenamefont {Jolie}, \citenamefont {Nomura},
  \citenamefont {Ahn}, \citenamefont {Cooper}, \citenamefont {Duckwitz},
  \citenamefont {Fitzler}, \citenamefont {Fransen}, \citenamefont {Gade},
  \citenamefont {Hinton}, \citenamefont {Ilie}, \citenamefont {Jessen},
  \citenamefont {Linnemann}, \citenamefont {Petkov}, \citenamefont
  {Pietralla},\ and\ \citenamefont {Radeck}}]{thomas2016}%
  \BibitemOpen
  \bibfield  {author} {\bibinfo {author} {\bibfnamefont {T.}~\bibnamefont
  {Thomas}}, \bibinfo {author} {\bibfnamefont {V.}~\bibnamefont {Werner}},
  \bibinfo {author} {\bibfnamefont {J.}~\bibnamefont {Jolie}}, \bibinfo
  {author} {\bibfnamefont {K.}~\bibnamefont {Nomura}}, \bibinfo {author}
  {\bibfnamefont {T.}~\bibnamefont {Ahn}}, \bibinfo {author} {\bibfnamefont
  {N.}~\bibnamefont {Cooper}}, \bibinfo {author} {\bibfnamefont
  {H.}~\bibnamefont {Duckwitz}}, \bibinfo {author} {\bibfnamefont
  {A.}~\bibnamefont {Fitzler}}, \bibinfo {author} {\bibfnamefont
  {C.}~\bibnamefont {Fransen}}, \bibinfo {author} {\bibfnamefont
  {A.}~\bibnamefont {Gade}}, \bibinfo {author} {\bibfnamefont {M.}~\bibnamefont
  {Hinton}}, \bibinfo {author} {\bibfnamefont {G.}~\bibnamefont {Ilie}},
  \bibinfo {author} {\bibfnamefont {K.}~\bibnamefont {Jessen}}, \bibinfo
  {author} {\bibfnamefont {A.}~\bibnamefont {Linnemann}}, \bibinfo {author}
  {\bibfnamefont {P.}~\bibnamefont {Petkov}}, \bibinfo {author} {\bibfnamefont
  {N.}~\bibnamefont {Pietralla}},\ and\ \bibinfo {author} {\bibfnamefont
  {D.}~\bibnamefont {Radeck}},\ }\href
  {https://doi.org/https://doi.org/10.1016/j.nuclphysa.2015.12.010} {\bibfield
  {journal} {\bibinfo  {journal} {Nucl. Phys. A}\ }\textbf {\bibinfo {volume}
  {947}},\ \bibinfo {pages} {203} (\bibinfo {year} {2016})}\BibitemShut
  {NoStop}%
\bibitem [{\citenamefont {Nomura}\ \emph {et~al.}(2016)\citenamefont {Nomura},
  \citenamefont {Rodr\'{\i}guez-Guzm\'an},\ and\ \citenamefont
  {Robledo}}]{nomura2016zr}%
  \BibitemOpen
  \bibfield  {author} {\bibinfo {author} {\bibfnamefont {K.}~\bibnamefont
  {Nomura}}, \bibinfo {author} {\bibfnamefont {R.}~\bibnamefont
  {Rodr\'{\i}guez-Guzm\'an}},\ and\ \bibinfo {author} {\bibfnamefont {L.~M.}\
  \bibnamefont {Robledo}},\ }\href {https://doi.org/10.1103/PhysRevC.94.044314}
  {\bibfield  {journal} {\bibinfo  {journal} {Phys. Rev. C}\ }\textbf {\bibinfo
  {volume} {94}},\ \bibinfo {pages} {044314} (\bibinfo {year}
  {2016})}\BibitemShut {NoStop}%
\bibitem [{\citenamefont {Garrett}\ \emph {et~al.}(2022)\citenamefont
  {Garrett}, \citenamefont {Zielińska},\ and\ \citenamefont
  {Clément}}]{garrett2022}%
  \BibitemOpen
  \bibfield  {author} {\bibinfo {author} {\bibfnamefont {P.~E.}\ \bibnamefont
  {Garrett}}, \bibinfo {author} {\bibfnamefont {M.}~\bibnamefont
  {Zielińska}},\ and\ \bibinfo {author} {\bibfnamefont {E.}~\bibnamefont
  {Clément}},\ }\href
  {https://doi.org/https://doi.org/10.1016/j.ppnp.2021.103931} {\bibfield
  {journal} {\bibinfo  {journal} {Prog. Part. Nucl. Phys.}\ }\textbf {\bibinfo
  {volume} {124}},\ \bibinfo {pages} {103931} (\bibinfo {year}
  {2022})}\BibitemShut {NoStop}%
\bibitem [{\citenamefont {Duval}\ and\ \citenamefont
  {Barrett}(1981)}]{duval1981}%
  \BibitemOpen
  \bibfield  {author} {\bibinfo {author} {\bibfnamefont {P.~D.}\ \bibnamefont
  {Duval}}\ and\ \bibinfo {author} {\bibfnamefont {B.~R.}\ \bibnamefont
  {Barrett}},\ }\href {https://doi.org/10.1016/0370-2693(81)90321-X} {\bibfield
   {journal} {\bibinfo  {journal} {Phys. Lett. B}\ }\textbf {\bibinfo {volume}
  {100}},\ \bibinfo {pages} {223} (\bibinfo {year} {1981})}\BibitemShut
  {NoStop}%
\bibitem [{\citenamefont {Nomura}\ \emph {et~al.}(2012)\citenamefont {Nomura},
  \citenamefont {Rodr\'{\i}guez-Guzm\'an}, \citenamefont {Robledo},\ and\
  \citenamefont {Shimizu}}]{nomura2012sc}%
  \BibitemOpen
  \bibfield  {author} {\bibinfo {author} {\bibfnamefont {K.}~\bibnamefont
  {Nomura}}, \bibinfo {author} {\bibfnamefont {R.}~\bibnamefont
  {Rodr\'{\i}guez-Guzm\'an}}, \bibinfo {author} {\bibfnamefont {L.~M.}\
  \bibnamefont {Robledo}},\ and\ \bibinfo {author} {\bibfnamefont
  {N.}~\bibnamefont {Shimizu}},\ }\href
  {https://doi.org/10.1103/PhysRevC.86.034322} {\bibfield  {journal} {\bibinfo
  {journal} {Phys. Rev. C}\ }\textbf {\bibinfo {volume} {86}},\ \bibinfo
  {pages} {034322} (\bibinfo {year} {2012})}\BibitemShut {NoStop}%
\end{thebibliography}%

\end{document}